\documentstyle[epsf,times]{mn}

\newcommand{\simgt}{\lower.5ex\hbox{$\; \buildrel > \over \sim \;$}}
\newcommand{\simlt}{\lower.5ex\hbox{$\; \buildrel < \over \sim \;$}}
\newcommand{\bm}[1]{\mbox{{\it \boldmath$#1$}}}

\newcommand{\skaco}[1]{\langle{#1}\rangle}

\newcommand{\apj}{ApJ}
\newcommand{\mnras}{MNRAS}
\newcommand{\LCDM}{$\Lambda$CDM~}

\newcommand{\rvir}{r_{\rm vir}}

\newcommand{\baredth}{\;\overline{\raise1.0pt\hbox{$'$}\hskip-6pt
\partial}\;}
\newcommand{\edth}{\;\raise1.0pt\hbox{$'$}\hskip-6pt\partial\;}

\begin{document}
\title[The Kurtosis of the Cosmic Shear Field]
{The Kurtosis of the Cosmic Shear Field}

\author[M. Takada \& B. Jain]
{Masahiro Takada \thanks{E-mail: mtakada@hep.upenn.edu}
and Bhuvnesh Jain \thanks{E-mail: bjain@physics.upenn.edu} \\
 Department of Physics
and Astronomy, University of Pennsylvania, 209 S. 33rd Street,
Philadelphia, PA 19104, USA
}

\onecolumn
\pagerange{\pageref{firstpage}--\pageref{lastpage}}

\maketitle

\label{firstpage}

\begin{abstract}
 We study the fourth-order moment of the cosmic shear field using the
 dark matter halo approach to describe the nonlinear gravitational
 evolution of structure in the universe. Since the third-order moment of
 the shear field vanishes because of symmetry, non-Gaussian signatures
 in its one-point statistics emerge at the fourth-order level.  We argue
 that the shear kurtosis parameter $S_{\gamma,4}\equiv
 \skaco{\gamma_i^4}_c/\skaco{\gamma_i^2}^3$ may be more directly
 applicable to realistic data than the well-studied higher-order
 statistics of the convergence field, since obtaining the convergence
 requires a non-local reconstruction from the measured shear field.  

 We compare our halo model predictions for the variance, skewness and
 kurtosis of lensing fields with ray-tracing simulations of cold dark
 matter models and find good agreement.  The shear kurtosis
 calculation is made tractable by developing approximations for fast and
 accurate evaluations of the 8-dimensional integrals necessary to obtain
 the shear kurtosis.  We show that on small angular scales,
 $\theta\simlt 5'$, more than half of the shear kurtosis arises from
 correlations within massive dark matter halos with $M\simgt
 10^{14}M_\odot$. The shear kurtosis is sensitive to the matter density
 parameter of the universe, $\Omega_{\rm m0}$, and has relatively weak
 dependences on other parameters.  Therefore, a detection of the shear
 kurtosis can be used to break degeneracies in determining $\Omega_{\rm
 m0}$ and the power spectrum amplitude $\sigma_8$ so far provided from
 measurements of the two-point shear statistics. 
The approximations we develop for the third- and fourth-order  moments allow 
for accurate halo model predictions for the 3-dimensional mass 
distribution as well. We demonstrate their accuracy 
in the small scale regime, below 2 Mpc, where analytical 
approaches used in the literature so far cease to be accurate. 

\end{abstract}

\section{Introduction}

Weak gravitational lensing caused by the large-scale
structure of the universe has been established as a 
useful probe of cosmological parameters
and offers the possibility of directly measuring the dark matter power
spectrum (see Mellier 1999 and Bartelmann \& Schneider 2001 for
reviews). Several independent groups have reported significant
detections of lensing by large-scale structure on distant
galaxy images ({\em cosmic shear}) from the ground
\cite{vW00,Wittman00,BRE00,Kaiser00,Maoli01,vW01a,Hoekstra02,Bacon02}
and from space \cite{Rhodes01,Haemm02,Ref02}.  These groups measured the
two-point correlation function of the cosmic shear field or the variance
of the filtered shear field and set constrains on cosmological
parameters, in particular some combination of the overall amplitude of
matter power spectrum ($\sigma_8$) and the matter density parameter of
the universe ($\Omega_{\rm m0}$), as shown in earlier theoretical work
\cite{Blandford91,Miralda91,Kaiser92,Villumsen96,BVM97,BhuvSelj97,Kaiser98}.

It has been shown that the non-Gaussian signature in the weak lensing
field induced by nonlinear gravitational clustering can be used to
break degeneracies in the determination of $\sigma_8$ and $\Omega_{\rm
m0}$ (Bernardeau, Van Waerbeke \& Mellier 1997; Jain \& Seljak
1997). This possibility is attractive, since it can determine
$\Omega_{\rm m0}$ via weak lensing measurements without invoking any
other methods such as the cosmic microwave background (CMB) and galaxy
redshift surveys. This also indicates that the dark energy component of
the universe can be constrained by combining lensing measurements with
the evidence for a flat universe revealed by the measured CMB angular
power spectrum (e.g., Netterfield et al. 2001).  Most theoretical work
so far has focused on the non-Gaussian signatures described by the
higher-order moments of the filtered convergence field (Bernardeau et
al. 1997; Jain \& Seljak 1997; Hui 1999; Van Waerbeke, Bernardeau \&
Mellier 1997; Jain, Seljak \& White 2000; Van Waerbeke et al. 2001b;
Munshi \& Jain 2001) or the skewness parameter in the aperture mass map
\cite{Schneider98,BS01}. It was recently also proposed that the genus
curve or Minkowski functionals of the convergence field could be an
efficient measure of the non-Gaussian signal
\cite{Sato01,Matsu01,Taruya02}.  Unfortunately these methods turn out to
have limitations in application to realistic data. Because realistic
data has a non-trivial survey geometry with many masked areas due to
light scattering, bright stars and so on, it is very challenging to
reconstruct the convergence from the measured shear field. On the other
hand, although the aperture mass method has the advantage of being
directly obtained from the shear map, it is likely to suffer from low
signal-to-noise ratio for the skewness measurement, because this method
uses a compensated filter \cite{Schneider96,Schneider98} and thus leads
to the loss of the non-Gaussian signal, especially on angular scales
smaller than a few arcminutes where the signal is large (see Van
Waerbeke et al. 2001a for detailed comparisons between various
two-point statistical measures of the shear field for actual data).

Very recently, Bernardeau, Van Waerbeke \& Mellier (2002a; hereafter
BvWM) proposed that some specific patterns in the three-point function
of the shear field can be used to extract the non-Gaussian
signal. Bernardeau, Mellier \& Van Waerbeke (2002b) then reported a
detection of this signal 
from actual data on $2-4$ arcminutes scales, although the
signal-to-noise so far is not enough to put robust constraints on $\Omega_{\rm
m0}$. The method proposed by BvWM appears to be a promising new measure
of non-Gaussianity. It is possible that their method loses some
non-Gaussian information because the vector-like property of their
statistic leads to partial cancellations between the signal on averaging. 
Their method also seems to have the limitation that
it cannot extract the signal on small scales ($\theta_{\rm s}\simlt
2'$), since their three-point function decreases for smaller 
scales and approaches zero at zero separation.

The purpose of this paper is to develop an alternative statistical
method directly applicable to the cosmic shear data. The method we
propose is the connected part of the fourth-order moment of the filtered
shear field, in particular the shear kurtosis parameter defined by
$S_{\gamma,4}=\skaco{\gamma_i^4}_c/\skaco{\gamma_i^2}^3$, since the
non-Gaussian signal appears first at the fourth-order level for the
one-point statistics. The kurtosis parameter collapses the information
in the trispectrum into a single less noisy quantity, although it does
not retain the full information in the four-point statistics.

Since the shear field on relevant angular scales is affected by the
nonlinear regime of structure formation (e.g., see Jain \& Seljak
1997), we need a model to correctly describe the redshift evolution and
statistical properties of gravitational clustering up to the four-point
level.  The perturbation theory well studied in the literature may not
be adequate for this task.  On the one hand, it is known that the
so-called `hyper-extended perturbation theory' \cite{Scocci99} can
describe the {\em strongly} nonlinear clustering regime (see for Hui
1999, van Waerbeke et al. 2001b and Munshi \& Jain 2001 for application
to weak lensing). However, the model does not describe the
intermediate-scale transition between the linear and strongly nonlinear
regimes \cite{CH01a,Scocci01}, which does affect weak lensing statistics
on a range of scales because of projection effects.  We therefore choose
to employ the dark matter halo approach, where gravitational clustering
is described in terms of correlations between and within dark matter
halos (see McClelland \& Silk 1977; Peebles 1980; Scherrer \&
Bertschinger 1991 for initial applications; for recent developments
see e.g. Sheth \& Jain 1997; Komatsu \& Kitayama 1999; Seljak 2000; Ma
\& Fry 2000; Scoccimarro et al. 2001; Cooray, Hu \& Miralda-Escude 2000;
Cooray \& Hu 2001a,b; and  Cooray \& Sheth 2002 for a recent review).  
There are several reasons
we use the halo model. First, the halo model is formally complete and
simple enough that higher-order statistics of the weak lensing fields
can be analytically calculated. Second, the results can be interpreted
in terms of halo properties, which is convenient for comparison with
other observations such as $X$-ray and optical surveys of clusters of
galaxies.  Finally, the model appears remarkably successful in that,
even though it relies on rather simplified assumptions, it has
reproduced results from numerical simulations
\cite{Seljak00,Ma00,Scocci01} and also allowed for interpretations of
observational results of galaxy clustering
\cite{Seljak00,Scocci01,Guzik02,Seljak02}.

Once the three ingredients of the halo model (halo profile, mass
function and halo bias) are specified, it is straightforward to develop
the formalism to calculate the shear kurtosis. Cooray \& Hu (2001a) have
investigated the bispectrum of the convergence field using the halo
model and find the convergence skewness is mainly due to rare and
massive halos on relevant scales of $\theta_{\rm s}\simlt 10'$, which is
referred to as the 1-halo term in this paper. We will also find that the
shear kurtosis arises mainly from the 1-halo term on the relevant
scales. However, since the direct application of the halo model requires
an 8-dimensional integration to obtain the 1-halo term, we develop an
approximation that significantly reduces the computational time and
gives the shear kurtosis with $10\%$ accuracy at most on angular scales
of interest.  Our model predictions will be compared in detail with
ray-tracing simulation results for all the statistical measures we
investigate: the convergence or shear variance, the convergence skewness
and the kurtosis parameters of the convergence and shear fields.
This comparison addresses the broader issue of whether the halo
model can accurately describe statistical properties of weak lensing
fields for higher-order moments beyond the two-point statistics
well studied in the literature.
We will pay special attention to the dependences of the shear
kurtosis on the cosmological parameters, $\Omega_{\rm m0}$ and
$\sigma_8$, for flat CDM (cold dark matter) models.

The outline of this paper is as follows. In \S \ref{halo} we present the
dark matter halo model used in this paper and then write down the
expressions for the power spectrum, bispectrum and trispectrum for the
underlying density field. In \S \ref{formul} we investigate the validity
of the halo model for weak lensing statistics by comparing the
predictions with the ray-tracing simulation results for the variance and 
skewness of the filtered convergence field. 
In \S \ref{app} we
develop an approximation for calculating the convergence kurtosis and
extend it to the shear kurtosis calculation in \S \ref{shearapp}.  
The dependence
of the shear kurtosis on cosmological parameters is presented in \S
\ref{results}. Finally, \S \ref{disc} is devoted to a summary and
discussion.  In the following, without explicit mention we will often
consider two CDM models: one is the SCDM model with
$\Omega_{\rm m0}=1$, $h=0.5$ and $\sigma_8=0.6$ and the other is the
$\Lambda$CDM model with $\Omega_{\rm m0}=0.3$, $\Omega_{\lambda0}=0.7$,
$h=0.5$ and $\sigma_8=0.9$, respectively. Here, $\Omega_{\rm m0}$ and
$\Omega_{\lambda0}$ are the present-day density parameters of matter and
cosmological constant, $h$ is the Hubble parameter, and $\sigma_8$ is
the rms mass fluctuations of a sphere of $8h^{-1}$Mpc radius.  The
choice of $\sigma_8$ for each model is motivated by the cluster
abundance analysis \cite{Eke96}.

\section{Dark matter halo approach}
\label{halo}

\subsection{Ingredients}
In the dark matter halo approach the underlying density field can be
described in terms of correlations between and within dark matter halos,
which are taken to be locally biased tracers of density perturbations in the
linear regime.  The method is therefore based on three essential
ingredients well studied in the literature: the mass function of dark
matter halos, the halo biasing function, and the halo density profile.

For the halo mass function, we adopt an analytical fitting model
proposed by Sheth \& Tormen (1999), which is more accurate on cluster
mass scales than the original Press-Schechter model \cite{PS74}. The
number density of halos with mass in the range between $M$ and $M+dM$ is
given by
\begin{eqnarray}
\frac{dn}{dM}dM&=&\frac{\bar{\rho}_0}{M}f(\nu)d\nu\nonumber\\
&=&\frac{\bar{\rho}_0}{M}A[1+(a\nu)^{-p}]\sqrt{a\nu}\exp\left(-\frac{a\nu}{2}
\right)\frac{d\nu}{\nu},
\label{eqn:massfun}
\end{eqnarray}
where $\nu$ is the peak height defined by
\begin{equation}
\nu=\left[\frac{\delta_c(z)}{D(z)\sigma(M)}\right]^2,
\end{equation}
$\bar{\rho}_0$ is the mean cosmic mass density today (we use
comoving coordinates throughout) and the numerical coefficients $a$ and
$p$ are empirically fitted from N-body simulations as $a=0.707$ and
$p=0.3$. The coefficient $A$ is set by the normalization condition
$\int_0^\infty\!d\nu f(\nu)=1$, leading to $A\approx 0.129$. Here
$\sigma(M)$ is the present-day rms fluctuations in the matter density
top-hat smoothed over a scale $R_M\equiv (3M/4\pi\bar{\rho}_0)^{1/3}$,
$D(z)$ is the growing factor (e.g., see Peebles 1980), and $\delta_c(z)$
is the threshold overdensity for spherical collapse model (see Nakamura
\& Suto 1997 and Henry 2000 for useful fitting functions). It should be
noted that the peak height $\nu$ is given as a function of $M$ at any
redshift.

Mo \& White (1996) developed a useful formula to describe the bias
relation between the dark matter halo distribution and the underlying
density field.  This idea has been improved by several authors using
N-body numerical simulations \cite{Mo97,Sheth99,ST99}; we will use the
fitting formula of Sheth \& Tormen (1999) for consistency with the mass
function (\ref{eqn:massfun}):
\begin{equation}
b(\nu)=1+\frac{a\nu-1}{\delta_c}+\frac{2p}{\delta_c(1+(a\nu)^p)},
\label{eqn:bias}
\end{equation}
where we have assumed scale-independent bias and neglected the
higher order bias functions $(b_2, b_3,\cdots)$ that have a negligible
effect on our final results. 
 
The density profile of dark matter halos is defined to be an average
over all halos with a given mass $M$ and does not necessarily assume all
halos have the same profile and spherical symmetry as stressed by Seljak
(2000). It is not evident that this argument should be valid for the
higher-order moments of the density field or the weak lensing
field. However, the agreement between our model predictions and
numerical simulations indicates that there is no strong violation of
the assumption.
Throughout this paper we assume the NFW model for the averaged halo
profile (Navarro, Frenk \& White 1996, 1997; hereafter NFW):
\begin{equation}
\rho(r)=\frac{fc^3M}{4\pi \rvir^3}\frac{1}{cr/\rvir (1+cr/\rvir)^2},
\label{eqn:nfw}
\end{equation}
where $f=1/[\ln(1+c)-c/(1+c)]$ and $\rvir$ is the virial radius of the
halo. The virial radius can be expressed in terms of the halo mass $M$
and redshift $z$ based on the spherical collapse model: $M=(4\pi
\rvir^3/3) \bar{\rho}_0\Delta(z)$, where $\Delta(z)$ is the overdensity
of collapse given as a function of redshift (e.g., see Nakamura \& Suto
1997 and Henry 2000 for a useful fitting formula). We have
$\Delta\approx 340$ for the \LCDM model.  It is worth noting that some
studies based on N-body simulations with higher resolution than in NFW
have suggested a steeper slope for the inner profile with $\rho\propto
r^{-1.5}$ at $r\simlt r_{\rm vir}/c$
\cite{Fukushige97,Moore98,JS00,Fukushige01}, whereas the predictions for the
outer parts of halos are in agreement with NFW: $\rho\propto r^{-3}$ at
$r\simgt r_{\rm vir}/c$. Lensing statistics on angular scales
of interest are affected more
strongly by the outer part of the density profile. Further the outer profile
is scaled by the concentration parameter $c$ for a given virial radius,
so we simply assume the NFW profile and pay close attention to 
the appropriate choice of $c$ as discussed below.

To give the halo profile in terms of $M$ and $z$, we further need to
express the concentration parameter $c$ in terms of $M$ and $z$;
however, this still remains somewhat uncertain.  The concentration $c$
is theoretically expected to be a weak function of halo mass as given by
$c=c_0 (M/M_\ast)^\beta$, where the normalization is $c_0\sim O(10)$ at
the present-day nonlinear mass scale $M_\ast$ defined by
$\delta_c(z=0)/\sigma(M_\ast)=1$ and the slope is $\beta \sim
-O(10^{-1})$.  We employ the form motivated by Seljak (2000):
\begin{equation}
c(M,z)=10(1+z)^{-1}\left(\frac{M}{M_{\ast}(z=0)}\right)^{-0.2}, 
\label{eqn:conc}
\end{equation}
where we have assumed the redshift dependence $(1+z)^{-1}$ as supported
by numerical simulations \cite{Bullock01}. There are several reasons we
adopt the form (\ref{eqn:conc}) for the unknown concentration
parameter. As for the slope $\beta$, we assume $\beta=-0.2$ which is
steeper than $\beta=-0.13$ originally proposed by NFW and Bullock et al.
(2001). This is motivated by the fact that for the NFW profile
(\ref{eqn:nfw}) the halo model with $\beta=-0.2$ can better reproduce
the well-studied nonlinear matter power spectrum than the model with
$\beta=-0.13$ as shown in Seljak (2000; also see Cooray et al. 2000).
As will be shown below, our model can also reproduce the simulation
results for the higher-order statistics of weak lensing fields on
relevant angular scales. In this sense, for the purpose of using the
halo model to describe the nonlinear gravitational evolution, it seems
to be appropriate to choose the halo model parameters so that the model
can reproduce the matter power spectrum as the first step.  The choice
of the normalization of $c_0=10$ at $M_\ast$ is supported by N-body
simulations (Bullock et al. 2001) and also validated by the fact that
the form (\ref{eqn:conc}) is consistent with recent observational
results of $c\sim O(10)$ on galactic scales of $M\sim 10^{12}M_\odot$
obtained from analyses of galaxy rotation curves \cite{Jimenez02} and
galaxy-galaxy lensing \cite{Guzik02,Seljak02}.  We will discuss in more
detail how possible variations in the concentration parameter affect the
final results of the shear kurtosis.

The normalized Fourier transform of the NFW profile (\ref{eqn:nfw}) is
given by
\begin{eqnarray}
y(k,M; z)=\frac{1}{M}\int^{\rvir}_0 4\pi r^2dr \rho(r)\frac{\sin kr}{kr},
\label{eqn:fnfw}
\end{eqnarray}
where $y(k)$ has the asymptotic behavior $y(k)\approx 1$ and
$y(k)\propto k^{-2}$ for $k \rvir/c\ll 1$ and $k\rvir/c\gg 1$,
respectively.

\subsection{The power spectrum, bispectrum and trispectrum}

The power spectrum $P(k)$, bispectrum $B(\bm{k}_1,\bm{k}_2,\bm{k}_3)$
and trispectrum $T(\bm{k}_1,\bm{k}_2,\bm{k}_3, \bm{k}_4)$ of the dark
matter density fluctuation are defined by
\begin{eqnarray}
\skaco{\delta(\bm{k}_1)\delta(\bm{k}_2)}&=&(2\pi)^3P(k_1)\delta_D(\bm{k}_{12}), 
\nonumber\\
\skaco{\delta(\bm{k}_1)\delta(\bm{k}_2)\delta(\bm{k}_3)}&=&
(2\pi)^3B(\bm{k}_1,\bm{k}_2,\bm{k}_3)\delta_D(\bm{k}_{123})\nonumber\\
\skaco{\delta(\bm{k}_1)\dots\delta(\bm{k}_{4})}_c&=&(2\pi)^3
T(\bm{k}_1,\bm{k}_2,\bm{k}_3,\bm{k}_4)\delta_D(\bm{k}_{1234}),
\end{eqnarray}
where $\bm{k}_{i\dots j}=\bm{k}_i+\dots+\bm{k}_j$, $\delta_D(\bm{k})$ is
the delta function, and $\skaco{\dots}$ denotes the ensemble average. 
The subscript $c$ denotes the connected part;
the trispectrum is identically zero for a
Gaussian field. 

In the picture of the halo approach, the power spectrum can be expressed
as the sum of correlations within a single halo (denoted the $1h$ term) 
and between different halos (the $2h$ term);
\begin{equation}
P(k)=P^{1h}(k)+P^{2h}(k),
\label{eqn:pshalo}
\end{equation}
with 
\begin{eqnarray}
P^{1h}(k)&=&I^0_2(k,k),\nonumber\\
P^{2h}(k)&=&\left[I^1_1(k)\right]^2P^{\rm L}(k). 
\end{eqnarray}
In the above equations have used the notation of Cooray \& Hu (2001a):
\begin{eqnarray}
I^\beta_\mu(k_1,\dots,k_{\mu})\equiv \int\!\!dM\frac{dn}{dM}
\left(\frac{M}{\bar{\rho}_0}\right)^\mu b_\beta(M)
y(k_1,M)\dots y(k_\mu, M). 
\end{eqnarray}
Note that we set $b_0=1$, $b_1=b$ given by equation (\ref{eqn:bias}) and
$b_i=0$ for $i\ge 2$.  The quantity $P^{L}(k)$ denotes the linear power
spectrum, and its redshift evolution is given by
$P^{L}(k,z)=D^2(z)P^L(k,z=0)$, although we will often omit $z$ in the
argument for simplicity. The requirement that on large scales
($k\rightarrow 0$ and $ y\sim 1$) the 2-halo contribution to the power
spectrum reduce to the linear power spectrum imposes the condition
$\int\!\!d\nu f(\nu)b(\nu)=1$, which is automatically satisfied by
equations (\ref{eqn:massfun}) and (\ref{eqn:bias}) within a few percents.

Similarly, the bispectrum can be expressed as sum of the 1-halo, 
2-halo and 3-halo contributions:
\begin{eqnarray}
B=B^{1h}+B^{2h}+B^{3h},
\label{eqn:bisp}
\end{eqnarray}
with 
\begin{eqnarray}
B^{1h}&=&I^0_3(k_1,k_2,k_3),\nonumber\\
B^{2h}&=&P^L(k_1)I^1_2(k_2,k_3)I^1_1(k_1)+{\rm 2 ~perm.},\nonumber\\
B^{3h}&=&B^{\rm pt}(\bm{k}_1,\bm{k}_2,\bm{k}_3)I^1_1(k_1)I^1_1(k_2)I^1_1(k_3),
\end{eqnarray}
where $B^{\rm pt}$ denotes the bispectrum calculated by perturbation
theory and the explicit expression is given in Appendix \ref{app:pert}.

Finally, the trispectrum arises from four contributions involving 
one to four halos \cite{CH01b}:
\begin{equation}
T=T^{1h}+(T^{2h}_{31}+T^{2h}_{22})+T^{3h}+T^{4h},
\label{eqn:trisp}
\end{equation}
with 
\begin{eqnarray}
T^{1h}&=&I^0_4(k_1,k_2,k_3,k_4),\label{eqn:trisp1h}\\
T^{2h}_{31}&=&P^L(k_1)I^1_3(k_2,k_3,k_4)I^1_1(k_1)+{\rm 3 ~perm.},
\label{eqn:trisp2h31} \\
T^{2h}_{22}&=&P^L(k_{12})I^1_2(k_1,k_2)I^1_2(k_3,k_4)+{\rm 2 ~perm.}, 
\label{eqn:trisp2h22} \\
T^{3h}&=&B^{\rm pt}(\bm{k}_1,\bm{k}_2,\bm{k}_3)I^1_2(k_3,k_4)
I^1_1(k_1)I^1_1(k_2)+{\rm 5~perm.}, \label{eqn:trisp3h}\\
T^{4h}&=&T^{\rm pt}(\bm{k}_1,\dots,\bm{k}_4)I_1^1(k_1)\dots I^1_1(k_4),
\label{eqn:trisp4h}
\end{eqnarray}
where $T^{\rm pt}$ denotes the trispectrum given by perturbation
theory (see Appendix \ref{app:pert}). Note that the 2-halo term is
further divided into two contributions, $T^{2h}_{31}$ and $T^{2h}_{22}$,
which represent taking three or two points in the first halo and then
one or two in the second halo.

\section{Validity of the halo model for weak lensing statistics}
\label{formul}

In this section, we investigate the validity of the halo model to
compute weak lensing statistics by comparing our model predictions to
ray-tracing simulations for the variance and skewness of the filtered
convergence field.

\subsection{Weak lensing convergence and shear fields}
The weak lensing convergence is expressed as a weighted projection of
the density
fluctuation field between source galaxy and the observer 
(e.g., see Mellier 1999 and Bartelmann \& Schneider 2001):
\begin{equation}
\kappa(\bm{\theta})=\int\!\!d\chi W(\chi,\chi_{\rm s}) 
\delta[\chi, d_A(\chi)\bm{\theta}],
\label{eqn:kappa}
\end{equation}
where $\chi$ is the comoving distance and the function $W$ is the
lensing weight function defined by
\begin{equation}
W(\chi,\chi_{\rm s})=\frac{3}{2}\Omega_{m0}H_0^2a^{-1}
\frac{d_A(\chi)d_A(\chi_{\rm s}-\chi)}{d_A(\chi_s)}. 
\label{eqn:weightgl}
\end{equation}
Here $H_0$ is the Hubble constant ($H_0=100h{~\rm km~s}^{-1}{\rm
Mpc}^{-1}$) and the function $d_A(\chi)$ is the comoving angular
diameter distance.  Note that throughout we assume all source galaxies
are at a single redshift $z_s$ for simplicity. The key simplification
used in equation (\ref{eqn:kappa}) is the Born approximation
\cite{Blandford91,Miralda91,Kaiser92}, where the convergence field is 
computed along the unperturbed path. Jain et al. (2000; hereafter JSW)
found that it is an excellent approximation for the two-point
statistics. Based on this result, we will assume that the Born
approximation also holds for the higher-order statistics we are
interested in.

A direct observable of weak lensing is the distortion effect on source
galaxy images characterized by the two components of the shear field,
$\gamma_1$ and $\gamma_2$, which correspond to elongations or
compressions along or at $45^\circ$ to $x$-axis, respectively.  In 
Fourier space, the shear fields $\gamma_1$ and $\gamma_2$ are simply
related to the convergence field via the relation
\begin{equation}
 \tilde{\gamma}_1(\bm{l})=\tilde{\kappa}(\bm{l})\cos(2\varphi_{l}),
\hspace{1em}
\tilde{\gamma}_2(\bm{l})=\tilde{\kappa}(\bm{l})\sin(2\varphi_{l}), 
\label{eqn:shear}
\end{equation}
where $\bm{l}=l(\cos\varphi_{l},\sin\varphi_{l})$, quantities with tilde
symbol denote their Fourier components and we have employed the flat-sky
approximation \cite{Blandford91,Miralda91,Kaiser92}.  Equation
(\ref{eqn:shear}) shows that $\gamma_{i}$ has a vector-like
property. More specifically, for example, each shear component could be
either positive or negative even around a dark matter halo on the sky,
whereas the convergence field is always positive.  The statistical
symmetry of the shear components around $0$ is the reason that all odd
moments of the shear field vanish. Hence the first non-vanishing
non-Gaussian signal appears at the fourth-order level for the one-point
statistics.

In practice spatially filtered lensing fields are used in order to
reduce the noise contribution due to the intrinsic ellipticities of
source galaxies.  The filtered shear field can be expressed as
\begin{equation}
\tilde{\gamma}^{F}_i(\bm{\theta}; \theta_{\rm s})=
\int\!\!\frac{d^2\bm{l}}{(2\pi)^2}\tilde{\gamma}_i(\bm{l})
F(l; \theta_{\rm s})\exp^{i\bm{l}\cdot\bm{\theta}}. 
\end{equation}
Throughout this paper, we employ the top-hat filter function with its 
Fourier transform given by
\begin{equation}
F(l;\theta_{\rm s})=2\frac{J_1(l\theta_{\rm s})}{l\theta_{\rm s}}, 
\label{eqn:tophat}
\end{equation}
where $J_1(x)$ is the first-order Bessel function. In the following, we
will omit the superscript ${\cal F}$ for the filtered fields of $\kappa$
and $\gamma$ for simplicity.
 
\subsection{Variance and higher-order moments of the 
filtered convergence field}

The variance of the filtered convergence field can be expressed as a
weighted integral of the dark matter power spectrum: 
\begin{eqnarray}
\sigma_\kappa^2(\theta_{\rm s})\equiv
\skaco{\kappa^2(\theta_{\rm s})}=\int \!\! d\chi W^2(\chi,\chi_s)
d_A^{-2}(\chi)
\int\!\!\frac{ldl}{2\pi}P\!\left(k=\frac{l}{d_A}; \chi
\right)F^2(l). 
\label{eqn:variconv}
\end{eqnarray}
This equation is derived by using the Limber approximation (Limber 1954;
also see Kaiser 1992) under the flat sky approximation. It should be
noted that the angular mode $l$ is related to the three dimensional
wavenumber as $k=l/d_A$.  By using the expression in equation
(\ref{eqn:pshalo}) for $P(k)$, we can compute the convergence variance
based on the halo model.

\begin{figure}
  \begin{center}
    \leavevmode\epsfxsize=8.4cm \epsfbox{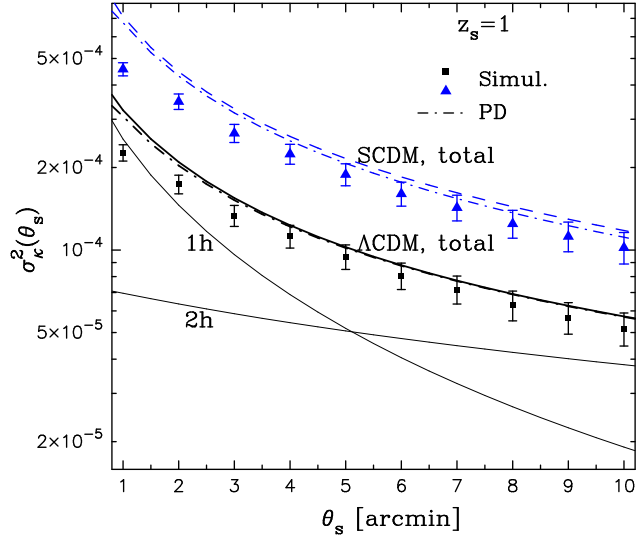}
  \end{center}
\caption{ The variance of the convergence field as a function of the
top-hat smoothing scale $\theta_{\rm s}$. The solid and dashed lines
show the halo model predictions for the SCDM and $\Lambda$CDM models,
respectively, with the source redshift $z_{\rm s}=1$.  The square and
triangle symbols are results from ray-tracing simulation, with error
bars giving the sample variance for a survey area of $25 ~{\rm
degree}^2$, calculated from the simulation data of Hamana \& Mellier
(2001).  The thin solid lines are the 1-halo and 2-halo contributions
for the $\Lambda$CDM model. The dot-dashed lines are the predictions
from the Peacock-Dodds fitting formula.  } \label{fig:comp_vari}
\end{figure}
\begin{figure}
  \begin{center}
    \leavevmode\epsfxsize=8.4cm \epsfbox{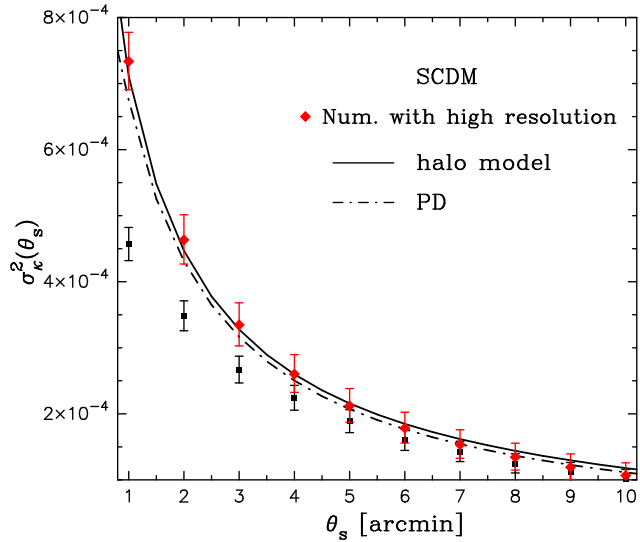}
  \end{center}
\caption{ Shown is the comparison of the high resolution simulation
results (diamond symbol) of $\sigma_\kappa^2$ (Jain et al.2000) with the
results (square) of lower resolution data (Hamana \& Mellier 2001) for
SCDM. The latter data is mainly used for the comparisons with model
predictions in this paper as explained in the text.  The solid and
dot-dashed lines are the predictions of the halo model and PD,
respectively, as in Figure \ref{fig:comp_vari}.  Note that the error
bars for the JSW data correspond to the sample variance for a survey
area of $2.8\times 2.8~{\rm degree}^2$.  }  \label{fig:comp_num}
\end{figure}
\begin{figure}
  \begin{center}
    \leavevmode\epsfxsize=8.cm \epsfbox{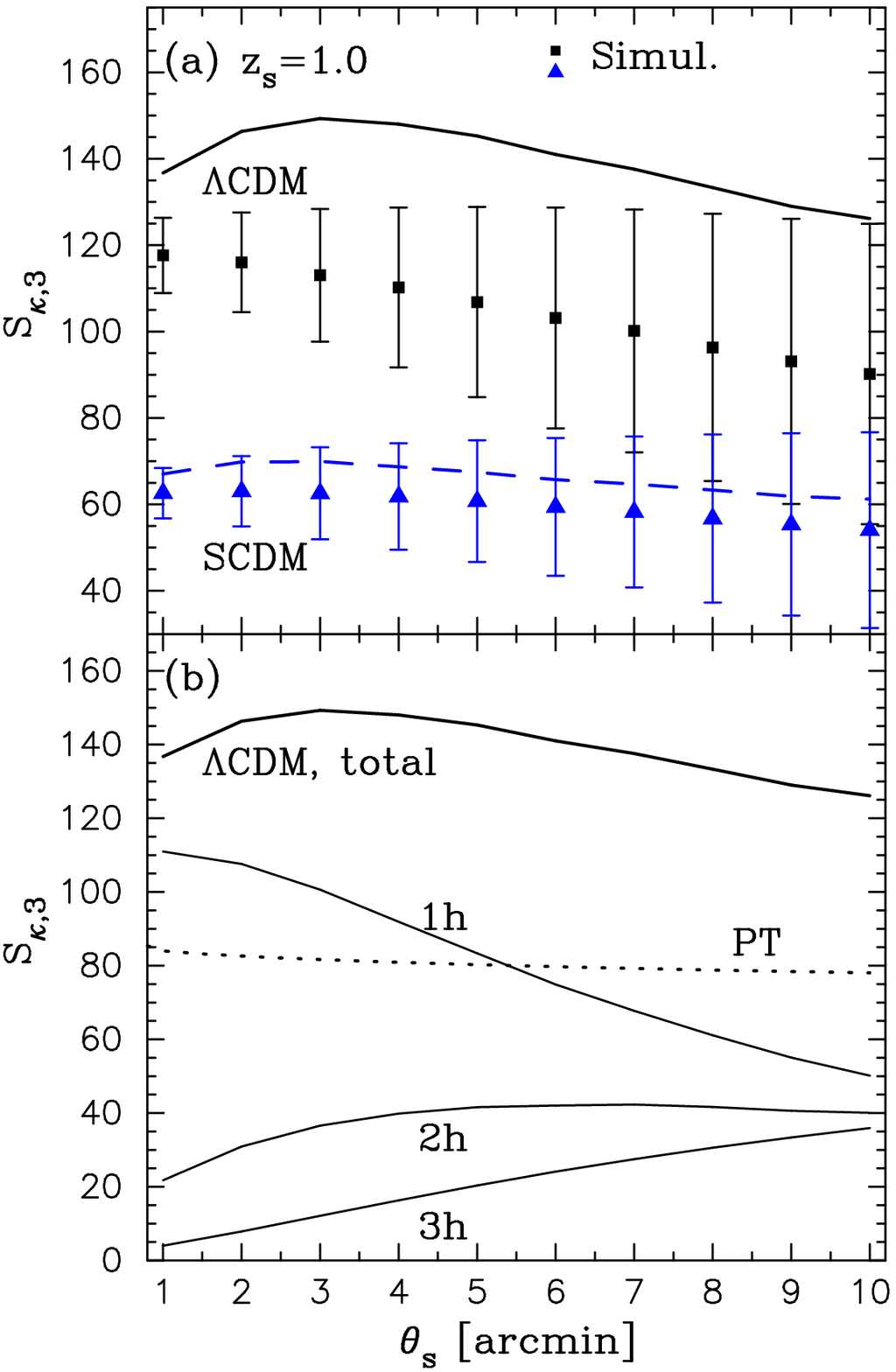}
  \end{center}
\caption{ The skewness parameter as a function of $\theta_{\rm s}$ as in
Figure \ref{fig:comp_vari}.  The upper panel shows a comparison of the
halo model predictions with the simulation results, while the lower
panel shows contributions from the 1-halo, 2-halo and 3-halo terms for
the $\Lambda$CDM model.  For comparison, the dotted line in the lower
panel shows the result predicted by the second-order perturbation
theory.  } \label{fig:comp_skew}
\end{figure}
Figure \ref{fig:comp_vari} plots the convergence variance as a function
of the top-hat smoothing scale for the SCDM ($\Omega_{\rm m0}=1$,
$h=0.5$, $\sigma_8=0.6$) and $\Lambda$CDM ($\Omega_{\rm m0}=0.3$,
$\Omega_{\lambda0}=0.7$, $h=0.7$, $\sigma_8=0.9$) models.  We fix
$z_{\rm s}=1$ for the source galaxy redshift.  For the linear matter
power spectrum used in the calculation, we employ a scale invariant
spectrum of the primordial fluctuations with the BBKS transfer function
\cite{BBKS86}.  The solid and dashed lines show the results of our halo
model for the $\Lambda$CDM and SCDM models, respectively, while the
dot-dashed lines are the predictions of using the Peacock \& Dodds
(1996; hereafter PD) fitting formula for the nonlinear power spectra.
The 1-halo and 2-halo contributions are shown by the thin solid lines
for $\Lambda$CDM, and one can see that the variance arises mainly from
the 1-halo term on angular scales of $\theta_{\rm s}\simlt 5'$, where
nonlinear structures play important role to the weak lensing statistics
(see, e.g.  Jain \& Seljak 1997).
The symbols with error bars are the ray-tracing simulation results,
where the error in each bin denotes the sample variance for a weak
lensing survey with an area of $25~{\rm degree}^2$. The ray-tracing
simulation builds on an N-body simulation based on the particle-mesh
(PM) code and has been kindly made available to us by T. Hamana (for
details see Hamana \& Mellier 2001; hereafter HM).

It is clear from Figure \ref{fig:comp_vari} that the halo model
predictions are in good agreements with the PD results as well as with
the simulation results for both the $\Lambda$CDM and SCDM models. We
have indeed confirmed that for all cosmological models we consider in
this paper the halo model can reproduce the PD results for
$\sigma_\kappa^2(\theta_{\rm s})$ within $\sim 5\%$ accuracy on angular
scales of interest.  This success at the two-point level is partly due
to our choice (\ref{eqn:conc}) of the concentration parameter for the
NFW profile.  However, there are slight
differences between the predictions and the numerical results on small
angular scales $\theta_{\rm s}\simlt 3'$. This is possibly due to the
lack of the numerical resolution of the ray-tracing data, because the
higher resolution simulation used by JSW yields more power on such small
scales, which gives a better match to the theoretical predictions, as
explicitly shown in Figure \ref{fig:comp_num} 
(see also discussions in Taruya et al. 2002 for the resolution of the HM
data).
We prefer to use the HM
data for comparison to our model predictions because we can use 40
realizations of simulation data with $25~{\rm degree}^2$ for each CDM
model in order to correctly estimate the sample variance.
Having an adequate number of the realizations is crucial to study the
higher-order moments especially on large angular scales, $\theta_{\rm
s}\simgt 5'$, since the higher moments are more sensitive to sample
variance.

In analogy with the second moment, the third-order moment of the
filtered convergence field can be expressed in terms of the bispectrum as
\begin{eqnarray}
\skaco{\kappa^3(\theta_{\rm s})}=\int\!\!d\chi W^3(\chi,\chi_{\rm s})
d_A^{-4}(\chi)
\int\!\!\frac{d^2\bm{l}_1}{(2\pi)^2}\frac{d^2\bm{l}_2}{(2\pi)^2}
B\left(\bm{k}_1,\bm{k}_2,-\bm{k}_{12}\right)F(l_1)F(l_2)F(l_{12}),
\label{eqn:defconv3}
\end{eqnarray}
where $\bm{k}_i=\bm{l}_i/d_A(\chi)$ and $B$ for the halo model is given
by equation (\ref{eqn:bisp}).  We explicitly write down the 1-halo
contribution to $\skaco{\kappa^3}$:
\begin{eqnarray}
\skaco{\kappa^3(\theta_{\rm s})}^{1h}=
\int\!\!d\chi W^3(\chi,\chi_{\rm s})
d_A^{-4}(\chi)\int\!\!dM\frac{dn}{dM}
\left(\frac{M}{\bar{\rho}_0}\right)^3
\int\!\!\frac{d^2\bm{l}_1}{(2\pi)^2}
\frac{d^2\bm{l}_2}{(2\pi)^2}y(l_1,M)y(l_2,M)y(l_{12},M)
F(l_1)F(l_2)F(l_{12}).
\label{eqn:conv3rd}
\end{eqnarray}
Although the Fourier transform of the NFW profile, $y(k,M)$, is given as
a function of the three-dimensional wavenumber $k$, we will often use
$l$ for the argument of $y(k,M)$ according to the relation of
$k=l/d_A(\chi)$ for simplicity.  To obtain $\skaco{\kappa^3(\theta_{\rm
s})}^{1h}$, we need to perform a 5-dimensional numerical integration,
since we can eliminate one angular integration using statistical
symmetry.  The convergence skewness parameter is defined by
\begin{equation}
S_{\kappa,3}(\theta_{\rm s})\equiv \frac{\skaco{\kappa^3(\theta_{\rm s})}}
{\sigma^4_\kappa(\theta_{\rm s})}.
\label{eqn:skewconv}
\end{equation}
This form is motivated by the fact that in perturbation theory both the
numerator and denominator in equation (\ref{eqn:skewconv}) scale as
$\sim \delta_1^4$, where $\delta_1$ is the linear solution for the
density fluctuation field.  Hence the skewness becomes almost
independent of the power spectrum normalization $\sigma_8$, giving
roughly a dependence as $S_{\kappa,3}\propto \Omega_{\rm m0}^{-1}$
through the dependences of the angular distances and the growth rate of
the fluctuations (Bernardeau et al. 1997).  In the results shown below,
we use the halo model self-consistently to compute
$\sigma_{\kappa}(\theta_{\rm s})$ in the denominator of
$S_{\kappa,3}$. Since our halo model can reproduce the PD results for
$\sigma_\kappa^2$ within $5\%$ accuracy, this does not significantly
affect our results for the skewness or kurtosis parameters.

Figure \ref{fig:comp_skew} plots the convergence skewness parameter as a
function of the smoothing scale as in Figure \ref{fig:comp_vari}. It is
clear from the upper panel that the halo model prediction agrees well
with the simulation result for SCDM over all scales. For the \LCDM model
our model slightly overestimates the simulation result on small scales.
Among the possible reasons for this discrepancy, one is that the HM
simulation result may underestimate the true value of $S_{\kappa,3}$ due
to a lack of numerical resolution as explained in Figure
\ref{fig:comp_num}.  As shown in the lower right panel of Figure 18 in
JSW, the high-resolution N-body simulations yields $S_{\kappa,3}\approx
140$ for $\Lambda$CDM on small angular scales, which gives a better
match to our halo model prediction. However, the precise value of
$S_{\kappa,3}$ for the $\Lambda$CDM model in numerical simulations is
still perhaps an open issue.  Independent ray-tracing simulation
performed by White \& Hu (2000; hereafter WH) indicate
$S_{\kappa,3}\approx 110$ around $\theta_{\rm s}=4'$.
We found that an important difference in the WH simulations is the values
of cosmological parameters, since they use $\sigma_8=1.2$ and $\Gamma=0.2$
for the \LCDM model, whereas JSW and HM used $\sigma_8=0.9$ and
$\Gamma=0.21$. For the cosmological models used in WH, our
halo model predicts a $\sim 15\%$ decrease of $S_{\kappa,3}$ at
$\theta_{\rm s}\le 5'$ compared with Figure \ref{fig:comp_skew} and the
resulting $S_{\kappa,3}$ is then marginally consistent with the result
shown in Figure 9 of WH on the angular scales we have considered. This
is probably due to the increase of $\sigma_8$ from $0.9$ to $1.2$, which
affects the skewness in a complex way since nonlinear contributions are
significant in both its numerator and denominator. Thus on these small
scales the expectation from perturbation theory that $S_{\kappa,3}$ is
independent of $\sigma_8$ is not exactly valid.  The halo model predicts
that the skewness and kurtosis of lensing fields slightly decrease with
increasing $\sigma_8$ as shown in Figure \ref{fig:skurt_cosmo} for the
shear kurtosis.  Another difference between the N-body simulation codes
used in HM, WH and JSW is that the JSW data is based on the adaptive
particle-particle/particle-mesh (AP$^3$M) N-body simulations (see
Jenkins et al. 1998 in more detail), while the HM and WH data are based
on the particle-mesh (PM) simulations.  The AP$^3$M method is expected
to achieve higher resolution than PM method for similar mesh resolution.
We were able to use a new high-resolution simulation performed with an
AP$^3$M code using $512^3$ particles \cite{Hamana02} to compute the
skewness. 
The new data gives $S_{\kappa,3}\approx 137$ for $\theta_{\rm s}=1'$,
which agrees with the halo model result in Figure \ref{fig:comp_skew},
but we also find $S_{\kappa,3}\approx 125$ for $\theta_{\rm s}=4'$, a
value higher than the simulation result shown in Figure
\ref{fig:comp_skew}, but still lower than the halo model prediction at
the 1-$\sigma$ level.
Finally, we note that $S_{\kappa,3}$ has a stronger dependence on
$\Omega_{\rm m0}$ with decreasing $\Omega_{\rm m0}$; for example, slight
decrease of $\Delta\Omega_{\rm m0}=-0.05$ leads to a significant change
of $\Delta S_{\kappa,3}\approx 21$ at $\theta_{\rm s}=1'$ for flat CDM
models with the $\Lambda$CDM model taken to be the fiducial model, while
the skewness for the SCDM model is almost unchanged with $\Delta
S_{\kappa,3}\approx 1$. These scalings are
roughly consistent with the expectation scaling $S_{\kappa,3}\propto
\Omega_{\rm m0}^{-1}$.  Hence, the discrepancy between our model and the
simulation result for the \LCDM model corresponds to a relatively small
change in $\Omega_{\rm m0}$.

\begin{figure}
  \begin{center}
    \leavevmode\epsfxsize=15.cm \epsfbox{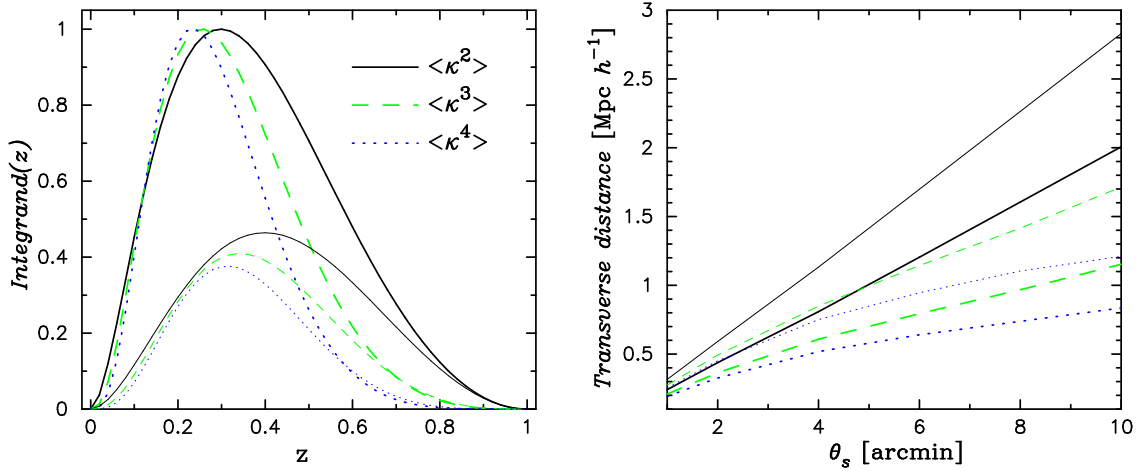}
  \end{center}
\caption{ The left panel shows the dependences of the integrand on
redshift for the variance (solid line) and the third- (dashed) and
forth-order (dotted) moments of the convergence field. The source
redshift and the filtering scale are $z_s=1$ and $\theta_{\rm s}=1'$.
The bold and thin lines are results for the SCDM and \LCDM models,
respectively, and each curve is normalized by the SCDM value at the peak
redshift. The figure shows that compared to the SCDM model, the
amplitude of each integrand decreases for $\Lambda$CDM and the peak
redshift shifts to higher $z$.  In the right panel, we show the comoving
transverse distance at the peak redshift of these integrands against the
angular smoothing scale.
} \label{fig:peakz_dist}
\end{figure}
In the lower panel of Figure \ref{fig:comp_skew}, the 1-halo, 2-halo and
3-halo contributions to $S_{\kappa,3}$ are separately plotted for
\LCDM. Notice that, for example, `1-halo' here means the convergence
third-order moment in $S_{\kappa,3}$ includes only the 1-halo
contribution, but the convergence variance used includes the total
contributions from the 1-halo and 2-halo terms.  It is apparent that the
1-halo term dominates over all scales shown and, in particular,
contributes $\sim 80\%$ of the total at the smallest scale $\theta=1'$.
This holds for the SCDM model also.  The result thus implies that the
higher-order moments are more sensitive to massive and rare dark matter
halos than the variance \cite{Scocci01,CH01a,CH01b}.  The dotted line
shows the skewness calculated by perturbation theory for the same
$\Lambda$CDM model, and it significantly underestimates $S_{\kappa,3}$,
since the weak lensing field on relevant scales is affected by strongly
nonlinear gravitational clustering. These features can be more
explicitly explained in Figure \ref{fig:peakz_dist}. In the left panel,
we plot how the integrand functions for the variance (solid line) and
the third- (dashed) and fourth-order (dotted) moments of the convergence
field depend on redshift for $z_s=1$ and $\theta_{\rm s}=1'$. Note that
the fourth-order moment is computed using the approximation developed
below.  The bold and thin lines are the results for the SCDM and \LCDM
models, respectively, where each curve is normalized by the SCDM value at
the peak redshift.  One can readily see that the higher-order moments
are more sensitive to lower redshift structures and, compared with the
result for SCDM, the amplitude of each integrand decreases for \LCDM and
the peak redshift shifts to higher $z$. The right panel plots the
comoving transverse distance at the peak redshift of the integrand
function as a function of the smoothing scale $\theta_{\rm s}$, which is
defined by $\lambda=d_A(z_{\rm peak})\theta_{\rm s}$.  Again, the figure
clarifies that the higher-order moments are more sensitive to structures
on smaller scales; for example, by comparing the solid and dotted lines
one finds that the transverse scales for the fourth-order moment are
smaller than those for the variance by factors of $0.8$ and $0.4$ at
$\theta_{\rm s}=1'$ and $10'$, respectively

It is worth noting differences between the convergence skewness, the
shear three-point correlation function recently proposed in BvWM (see
also Bernardeau, Mellier \& Van Waerbeke 2002), and the shear kurtosis.
As shown in Figure \ref{fig:comp_skew}, $S_{\kappa,3}$ has a weak
dependence on the angular scales as pointed out based on the
perturbation theory (Bernardeau et al. 1997), while the shear
three-point correlation function has a logarithmically decreasing
behavior with decreasing the angular scale as shown in Figure 6 in
BvWM. It is likely that their shear three-point function loses useful
non-Gaussian information resulting from cancellations between signals
caused by the vector-like property of the shear field.  An advantage of
the shear kurtosis parameter is that it collapses information from the
4-point statistics into a single quantity without being affected by such
cancellations.  However the kurtosis parameter is a higher-order moment,
and so it remains to be seen how its signal-to-noise properties compare
with the three-point function of BvWM.

\section{The kurtosis of the convergence field}
\label{app} 

In this section, we develop a useful approximation for fast and accurate
evaluations of the convergence kurtosis parameter. In particular, we
concentrate on developing approximations for calculating the 1-halo
term, $\skaco{\kappa^4(\theta_{\rm s})}^{1h}_c$, which provides the
dominant contribution to the convergence kurtosis on small angular
scales. The approximations for the 2-halo and 3-halo terms are presented
in Appendix \ref{conv23h}.  Those approximations will be used to develop
a method to compute the shear kurtosis in the next section.

\subsection{Definition}
The connected part of the fourth-order moment of the convergence field
is given by
\begin{eqnarray}
\skaco{\kappa^4(\theta_{\rm s})}_c=\int\!\!d\chi W^4(\chi,\chi_s)d_A^{-6}
\int\!\!\frac{d^2\bm{l}_1}{(2\pi)^2}\frac{d^2\bm{l}_2}{(2\pi)^2}
\frac{d^2\bm{l}_3}{(2\pi)^2}T(\bm{k}_1,\bm{k}_2,\bm{k}_3,-\bm{k}_{123})
F(l_1)F(l_2)F(l_3)F(l_{123}),
\label{eqn:fullconv4th} 
\end{eqnarray}
where $\bm{k}_i=\bm{l}_i/d_A(\chi)$ and the trispectrum is given by
equation (\ref{eqn:trisp}) within the framework of the halo model.  From
the results of the convergence skewness, it is expected that the most
important contribution to the fourth-order moment is the
1-halo term on angular scales of interest. The 1-halo term is given by
\begin{eqnarray}
\skaco{\kappa^4(\theta_{\rm s})}_c^{1h}&=&\int\!\!d\chi W^4(\chi)d_A^{-6}
\int\!\!dM \frac{dn}{dM}\left(\frac{M}{\bar{\rho}_0}\right)^4
\int\!\!\frac{d^2\bm{l}_1}{(2\pi)^2}\frac{d^2\bm{l}_2}{(2\pi)^2}
\frac{d^2\bm{l}_3}{(2\pi)^2}y(l_1,M)y(l_2,M)
y(l_3,M)y(l_{123},M) \nonumber\\
&&\hspace{26em}\times F(l_1)F(l_2)F(l_3)F(l_{123}).
\label{eqn:conv4th}
\end{eqnarray}
Hence, to obtain $\skaco{\kappa^4(\theta_{\rm s})}^{1h}_c$, we have to
perform at least a 7 dimensional numerical integration, even after we
eliminate one angular integration of $\bm{l}_i$ using statistical
symmetry.  Direct integration is not suitable for our final purpose of
evaluating the dependence on cosmological parameters, which requires
lots of computations in parameter space.  We therefore explore an
approximation for calculating $\skaco{\kappa^4}_c$ with adequate
accuracy and reasonable computational expense.

Motivated by perturbation theory, as for the skewness
parameter, we consider the convergence kurtosis parameter defined by
\begin{equation}
S_{\kappa,4}(\theta_{\rm s})\equiv \frac{\skaco{\kappa^4(\theta_{\rm s})}_c}
{\sigma_\kappa^6(\theta_{\rm s})}.
\label{eqn:kurt} 
\end{equation}
We use self-consistently the halo model to calculate
$\sigma_{\kappa}(\theta_{\rm s})$ in the denominator of
$S_{\kappa,4}$. It is expected that $S_{\kappa,4}$ has a dependence
roughly given by perturbation theory as $S_{\kappa,4}\propto \Omega_{\rm
m0}^{-2}$.

The approximation for calculating $S^{1h}_{\kappa,4}$ developed below
allows us to simplify the three-dimensional angular integrations of
$d^2\bm{l}_1d^2\bm{l}_2d^2\bm{l}_3$ in equation (\ref{eqn:conv4th}),
whereby we can obtain $\skaco{\kappa^4(\theta_{\rm s})}^{1h}_c$ by a
5-dimensional numerical integration instead of the original
8-dimensional one.

\subsection{Quadrilateral configuration dependence}

Equation (\ref{eqn:conv4th}) shows that, although the integrand function
of $\skaco{\kappa^4(\theta_{\rm s})}^{1h}_c$ does depend on
quadrilateral configuration with four sides $\bm{l}_1$, $\bm{l}_2$,
$\bm{l}_3$ and $\bm{l}_4(=-\bm{l}_{123})$ in Fourier space, the angular
dependences of $\bm{l}_i$ appear only via $\bm{l}_{123}$ in
$y(l_{123},M)$ and $F(l_{123})$ as a result of the spherical symmetry of
the NFW profile\footnote{On the other hand, when
$\skaco{\kappa^4(\theta_{\rm s})}_c$ is calculated in perturbation
theory, the angular dependences appear via products of the $\bm{l}_i$
vectors in the perturbation trispectrum in addition to via
$\bm{l}_{123}$ in $y(l_{123})$ and $F(l_{123})$, as explicitly shown in
equation (\ref{eqn:perttrisp}).}. Because of statistical symmetry,
without loss of generality we can express any configurations in terms of
5 parameters; three side lengths of $l_1$, $l_2$ and $l_3$ and two
angles $\Phi_2$ and $\Phi_3$, where $\Phi_2$ is angle between $l_1$ and
$l_2$ and $\Phi_3$ angle between $l_{12}$ and $l_3$.  The side length
$l_{123}$ can then be expressed as
\begin{equation}
l_{123}=\sqrt{l_{12}^2+l_3^2-2l_{12}l_3\cos\Phi_3},
\label{eqn:l123}
\end{equation}
with $l_{12}=(l_1^2+l_2^2-2l_1l_2\cos\Phi_2)^{1/2}$. Note that the
volume element (\ref{eqn:conv4th}) of integration can be rewritten, 
after performing one of the angular integrals, as
$d^2\bm{l}_1d^2\bm{l}_2d^2\bm{l}_3=(2\pi)l_1dl_1l_2dl_2
l_3dl_3d\Phi_2d\Phi_3$.

\subsection{Approximation for the integration of the 
top-hat filter function}

First, we consider an approximation for the integration of the top-hat
filter function (\ref{eqn:tophat}) motivated by Appendix A in Bernardeau
(1994), where the geometrical properties of the integration of products
of the {\em three-dimensional} top-hat window function are derived.  In
Appendix \ref{app:tophat} we prove the following identity for the
integration of products of top-hat kernels:
\begin{eqnarray}
&&\int\!\!\frac{d^2\bm{l}_1}{(2\pi)^2}\frac{d^2\bm{l}_2}{(2\pi)^2}
\frac{d^2\bm{l}_3}{(2\pi)^2}F(l_1)F(l_2)F(l_3)F(l_{123})
=\int\!\!\prod_{i=1}^3\frac{l_idl_i}{2\pi}F^2(l_i).
\end{eqnarray}
The result above cannot be applied exactly to simplify equation
(\ref{eqn:conv4th}) because of the $y(l_{123})$ term. We therefore use
the following replacement for the filter function $F(l_{123})$ in
equation (\ref{eqn:conv4th}) as an approximation to be tested:
\begin{equation}
F(l_{123})\approx F(l_1)F(l_2)F(l_3). 
\label{eqn:apptop}
\end{equation}
The corresponding approximation for the three-dimensional window
function is used in Scoccimarro et al. (2001) for the study of the
skewness and kurtosis parameters of the three-dimensional density
field. It is worth noting that this approximation indeed becomes exact
if $y(l,M)=\rm{constant}$ in equation (\ref{eqn:conv4th}).  Hence, to
the extent that the regime $y(l,M)\approx 1$ at $l\simlt
d_A(\chi)c/r_{\rm vir}$ provides the main contribution to
$\skaco{\kappa^4(\theta_{\rm s})}_c^{1h}$ for a given $M$ and $z$, it is
reasonable to expect that the replacement (\ref{eqn:apptop}) is a good
approximation for realistic density profiles.

\subsection{Approximation for the convergence skewness}
\label{appconvskew}

Given the approximation (\ref{eqn:apptop}), the next problem we consider is
to explore an approximation to describe the configuration
dependence of $y(l_{123},M)$  in a way that allows us to evaluate the angular
integrations with respect to $\Phi_2$ and $\Phi_3$ in equation
(\ref{eqn:conv4th}).

For this purpose, let us begin by considering an approximation for
calculating the 1-halo term in the convergence third-order moment,
$\skaco{\kappa^3}^{1h}$. This is because the accuracy of our
approximation for $\skaco{\kappa^3}^{1h}$ can be tested by comparing the
prediction with the true value obtained by the direct integration, and
then it can be extended to the calculation of the fourth-order moment.
The dependence on the triangle configuration appears via $l_{12}$ in
$y(l_{12},M)$ with $l_{12}=(l_1^2+l_2^2-2l_1l_2\cos\Phi_2)^{1/2}$.  We
propose a method to expand $y(l_{12})$ around a fiducial triangle
configuration with a fixed $\Phi_2$ in analogy with the Taylor expansion
of $y(l_{12})$ with respect to $\Phi_2$, whereby we can analytically
perform the
angular integrations of $\bm{l}_i$ in equation (\ref{eqn:conv3rd}).  The
critical question that arises is: which fiducial configuration is
appropriate for the expansion?  This can be answered by using the halo
model analysis of Cooray \& Hu (2001a) for the convergence bispectrum,
which is part of the integrand of $\skaco{\kappa^3}$.  Figure 7 in their
paper explicitly illustrates the configuration dependence of the
bispectrum and implies that the main contribution to
$\skaco{\kappa^3(\theta_{\rm s})}$ arises from equilateral triangle
configurations with $l_1=l_2=l_{12}$.
Hence, it will be reasonable to take a prescription that the fiducial
configuration contains equilateral triangle configurations when
$l_1=l_2$. This holds for $\Phi_2=\pi/3$.  We thus propose the
following approximation for calculating $\skaco{\kappa^3(\theta_{\rm
s})}^{1h}$ combined with the approximation (\ref{eqn:apptop}):
\begin{eqnarray} 
\skaco{\kappa^3(\theta_{\rm s})}^{1h}\approx  
\int\!\!
d\chi W^3(\chi,\chi_{\rm s}) d_A^{-4}\int\!\!dM\frac{dn}{dM}
\left(\frac{M}{\bar{\rho}_0}\right)^3
\int\!\!\frac{l_1dl_1}{2\pi}
\frac{l_2dl_2}{2\pi}y(l_1)y(l_2)y(\tilde{l}_{12})
F^2(l_1)F^2(l_2),
\label{eqn:skewapp}
\end{eqnarray}
with $\tilde{l}_{12}=(l_1^2+l_2^2-l_1l_2)^{1/2}$. Note that
$\tilde{l}_{12}=l_1=l_2$ when $l_1=l_2$, so that the dimension of
integration is reduced from 5 to 4 in equation (\ref{eqn:conv3rd}). Like
the Taylor expansion, one can include higher-order corrections arising
from the expansion of $y(l_{12},M)$ at the order of
$O(\Phi_2-\pi/3)$\footnote{In this case, the expansion parameter
$(\Phi_2-\pi/3)$ could be larger than unity in the range of
$\Phi_2=[0,2\pi]$, so the convergence of the expansion is no longer
guaranteed.}.
We find that the zeroth-order
approximation (\ref{eqn:skewapp}) works remarkably well as shown
below.
\begin{figure}
  \begin{center}
    \leavevmode\epsfxsize=8.4cm \epsfbox{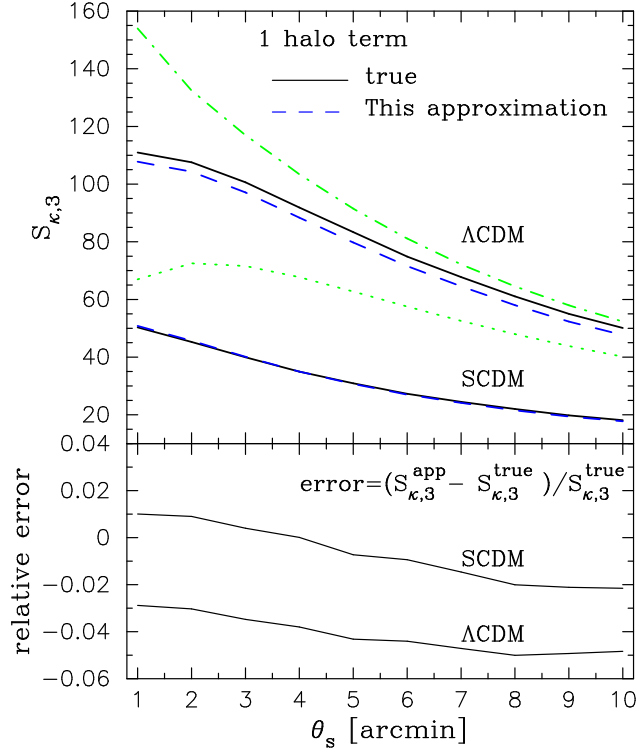}
  \end{center}
\caption{ Comparison of our approximation for the 1-halo term of
 convergence skewness parameter with direct integration values for the
 SCDM and $\Lambda$CDM models.  The solid lines are the direct
 integration values, while the dashed lines show the results of
 approximation (\ref{eqn:skewapp}).  For comparison, we also show the
 results calculated using other approximations for $\Lambda$CDM: the dot-dashed
 and dotted lines are computed using the replacements of
 $y(l_{12},M)\approx y(l_1,M)$ and $y(l_{12})\approx y(l_1,M)y(l_2,M)$,
 respectively, for the integration.  In the lower panel, the relative
 errors of our approximation are shown for the two models.}  \label{fig:app_skew}
\end{figure}

Figure \ref{fig:app_skew} demonstrates the accuracy of our approximation
(\ref{eqn:skewapp}) for the 1-halo term of the convergence skewness by
comparing the predictions with the direct integration results of
equation (\ref{eqn:conv3rd}) for the SCDM and \LCDM models.  The
approximation is very accurate, as its relative accuracy is better than
$5\%$ over all angular scales for both models.  For comparison, the
dotted and dot-dashed lines show the results of using other possible
approximations for $\Lambda$CDM, where we used the replacements of
$y(l_{12},M)= y(l_1,M)y(l_2,M)$ or $y(l_{12},M)= y(l_1,M)$ in equation
(\ref{eqn:conv3rd}), respectively, in addition to the approximation
(\ref{eqn:apptop}) for the filter function.  The former approximation is
motivated on the analogy of equation (\ref{eqn:apptop}), while
the latter is
indeed used by Scoccimarro et al. (2001) for calculations of the
skewness and kurtosis parameters of the three-dimensional density
field. It is clear that the approximation of $y(l_{12})=y(l_1)$
overestimates the value of $S_{\kappa,3}^{1h}$ (see also Cooray \& Hu
2001a) and the discrepancy is larger on smaller scales. In more detail,
it overestimates $S_{\kappa,3}^{1h}$ by $\sim 40\%$ at $\theta_{\rm
s}=1'$.  The approximation $y(l_{12},M)=y(l_1,M)y(l_2,M)$ underestimates
the skewness, since $y(l,M)\le 1$, and yields $S_{\kappa,3}^{1h}$
smaller by $\sim 60\%$ than the correct value. Hence, we cannot use
these approximations to predict the higher-order moments of weak lensing
fields with sufficient accuracy for our purpose.

\begin{figure}
  \begin{center}
    \leavevmode\epsfxsize=8.4cm \epsfbox{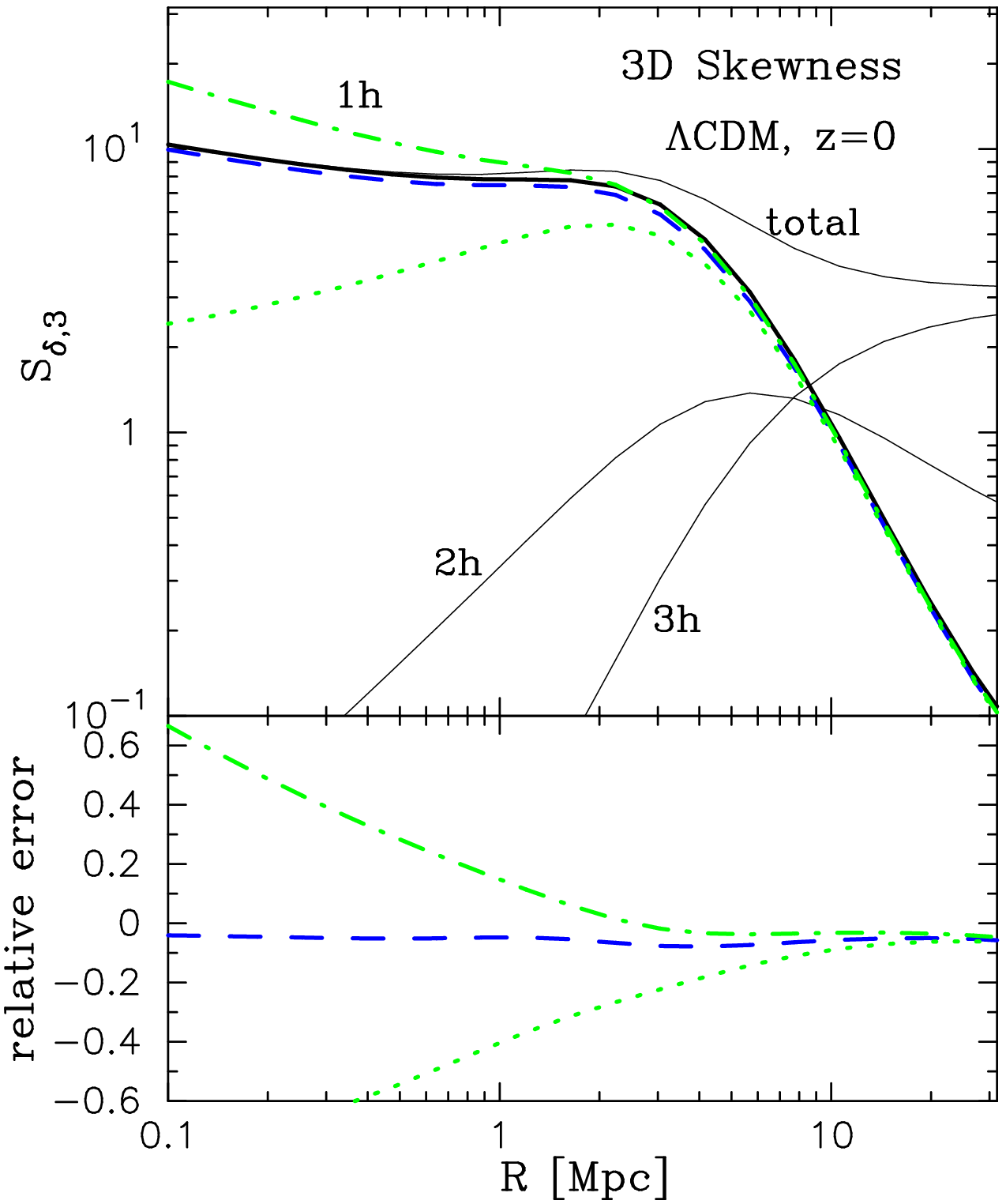}
  \end{center}
\caption{ The skewness parameter of the three-dimensional density field,
 $S_{\delta,3}=\skaco{\delta^3}/\skaco{\delta^2}^2$, as a
function of the top-hat smoothing scale, $R$ (Mpc), for the \LCDM model
at $z=0$. We demonstrate the performance of our approximation for the
1-halo contribution as shown in Figure \ref{fig:app_skew}. The dotted 
and dot-dashed lines show other approximations for the 1-halo term used in the
figure, as in Figure \ref{fig:app_skew}. The thin
lines are the 2-halo, 3-halo and total contributions. } \label{fig:3dskew}
\end{figure}
It is straightforward to apply our approximation to evaluations of the
skewness parameter of the three-dimensional density field,
$S_{\delta,3}\equiv \skaco{\delta^3}/\skaco{\delta^2}^2$, which is
relevant for surveys of galaxies clustering (e.g., see Scoccimarro et
al. 2001).  Figure \ref{fig:3dskew} plots the result against the
three-dimensional smoothing scale $R$ (Mpc) for the \LCDM model and
$z=0$ as shown in Figure \ref{fig:app_skew}. Note that we have used the
three-dimensional top-hat filter function, and equation
(\ref{eqn:apptop}) can be used as an approximation for the kernel.  It
is clear that our approximation again works well, implying that it
will be useful for efficiently exploring parameter space for
constraining cosmological parameters from $S_{\delta,3}$ measurements
down to very small scales. The issue of the small scale behavior of
higher-order moments is somewhat an open question since results 
from numerical simulations are not yet reliable for scales below 1 Mpc. 
The results shown in this paper for the third and fourth-order moment are 
encouraging. To the extent that the current halo model
describes clustering accurately, we have tractable analytical means of
predicting higher-order clustering statistics extending to very small scales. 

\subsection{Approximation for the convergence kurtosis}
\label{appconv}

\begin{figure}
  \begin{center}
    \leavevmode\epsfxsize=5.cm \epsfbox{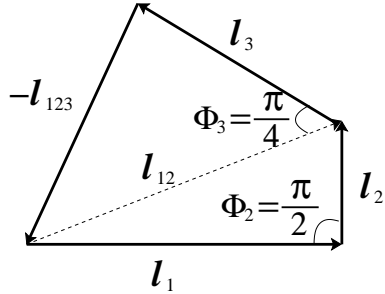}
  \end{center}
\caption{The sketch of the fiducial 4-point configuration used in the
approximation (\ref{eqn:appconv4th}) for calculation the convergence
fourth-order moment. The two angles $\Phi_2$ and $\Phi_3$ are set to be
$\pi/2$ and $\pi/4$, respectively, but the side length parameters $l_1$,
$l_2$ and $l_3$ are treated as variables.  
} \label{fig:4ptconf}
\end{figure}
\begin{figure}
  \begin{center}
    \leavevmode\epsfxsize=8.cm \epsfbox{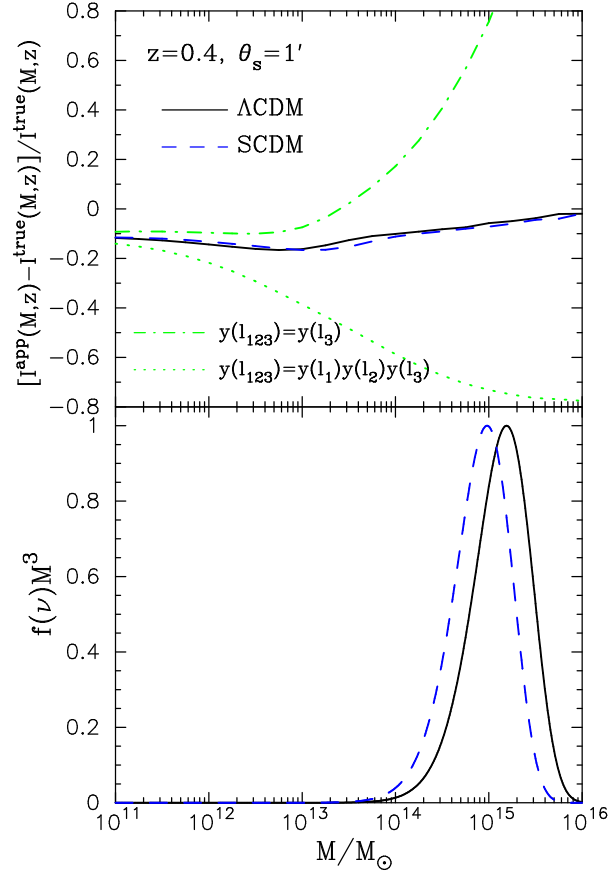}
  \end{center}
\caption{ Shown is the accuracy of our approximation
(\ref{eqn:kurtinteg}) for the integrand function ${\cal I}(M,z)$ for the
1-halo term of the convergence fourth-order moment. For the $\Lambda$CDM
and SCDM models, the relative errors defined by $({\cal I}^{\rm
app}-{\cal I}^{\rm true})/{\cal I}^{\rm true}$ are plotted as the solid
and dashed lines, respectively, as a function of halo mass $M$. We here
fix $z=0.4$ and $\theta_{\rm s}=1'$ for the lens redshift and the
smoothing scale.  For comparison, the dotted and dot-dashed lines show
the results of using other approximations as in Figure
\ref{fig:app_skew} (see text for more details).  In the lower panel, we
plot the mass function weighing, $f(\nu)M^3$, in the 1-halo term for the
models, where each curve is normalized to give unity at the peak scale.
} \label{fig:app_test}
\end{figure}
\begin{figure}
  \begin{center}
    \leavevmode\epsfxsize=8.cm \epsfbox{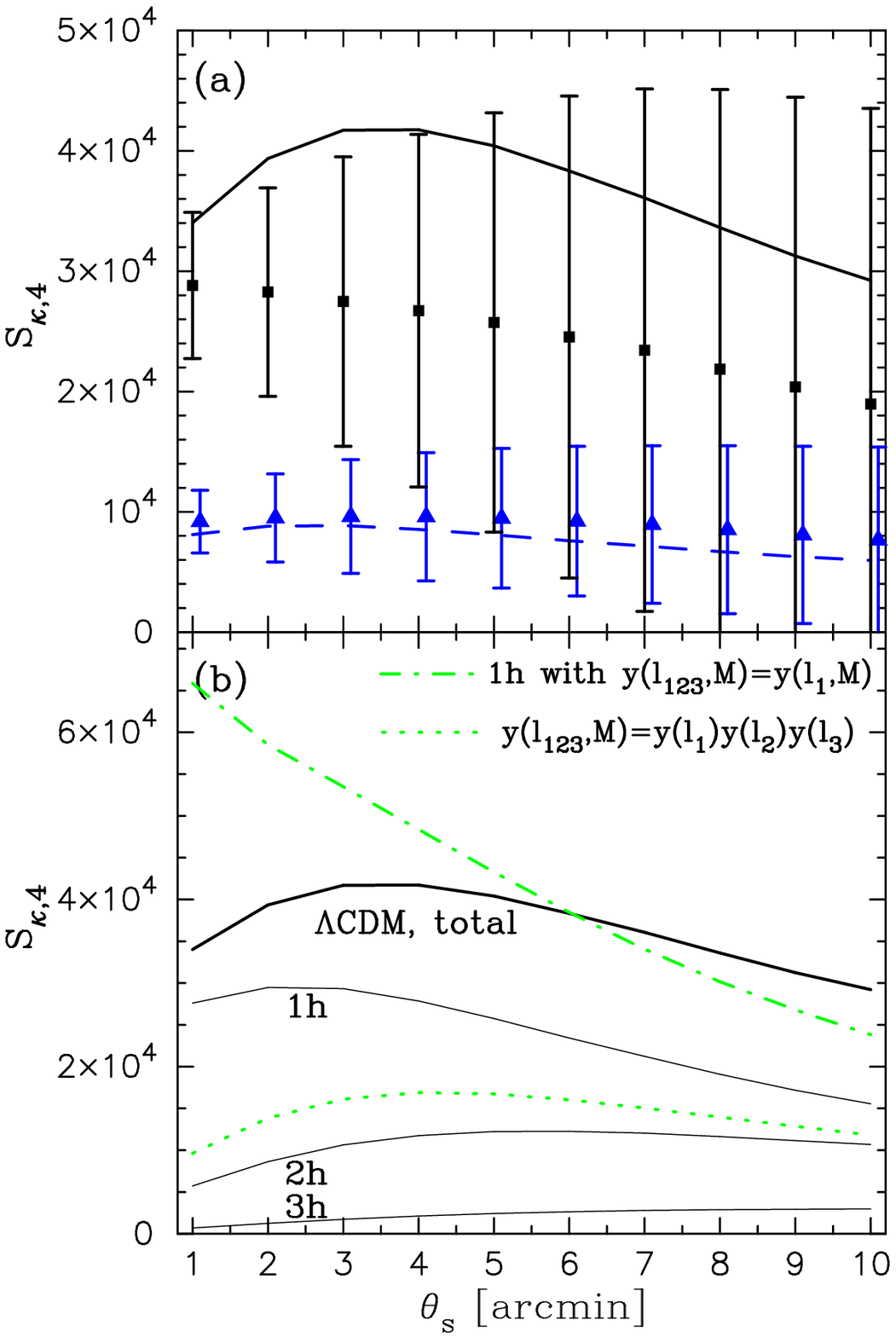}
  \end{center}
\caption{ Convergence kurtosis $S_{\kappa,4}$ as a function of the
smoothing scale as in Figure \ref{fig:comp_skew}.  The square and triangle
symbols with error bars are the simulation results for the $\Lambda$CDM
and SCDM models, respectively, while the solid and dashed lines denote
our model predictions.  For illustration, we slightly shift the
simulation result for SCDM along the x-axis.  The thin solid lines in
the lower panel are the 1-halo, 2-halo and 3-halo contributions for the
$\Lambda$CDM model.  For comparison, the dotted and dot-dashed lines are
the results of other approximations for the 1-halo term as in Figure
\ref{fig:app_test}. } \label{fig:comp_kurt}
\end{figure}

Based on the above success of our approximation for the skewness
parameter, we extend it to develop an approximation for the 1-halo term
of the convergence fourth-order moment, $\skaco{\kappa^4}^{1h}_c$. The
problem is to consider an efficient expansion of $y(l_{123},M)$ in
equation (\ref{eqn:conv4th}) with respect to the two angles $\Phi_2$ and
$\Phi_3$, such that we can analytically perform the two-dimensional angular
integrations of $\Phi_2$ and $\Phi_3$.
In the same spirit as the approximation (\ref{eqn:skewapp}), we choose
the fiducial configuration with $l_1=l_2=l_3=l_{123}$, 
because we believe, in analogy with the skewness, that the trispectrum
with such configurations produces the main contribution to the
fourth-order moment.  For the kurtosis, we need to make additional
choices for the angles $\Phi_{2}$ and $\Phi_{3}$; we simply set
$\Phi_{2}=\pi/2$ and $\Phi_3=\pi/4$, which implies a square shaped
configuration when $l_1=l_2=l_3$.
The sketch in Figure \ref{fig:4ptconf} illustrates the fiducial 4-point
configuration. 
Applying the above approximation to equation (\ref{eqn:conv4th}) gives
\begin{eqnarray}
\skaco{\kappa^4(\theta_{\rm s})}_c^{1h}\approx \int\!\!d\chi
W^4(\chi,\chi_{\rm s})d_A^{-6} 
\int\!\!dM\frac{dn}{dM} \left(\frac{M}{\bar{\rho}_0}\right)^4{\cal
I}(M,z; \theta_{\rm s}), \label{eqn:appconv4th}
\end{eqnarray}
with 
\begin{eqnarray}
{\cal I}(M,z; \theta_{\rm s})\equiv 
\int\!\!\prod_{i=1}^3\frac{l_idl_i}{2\pi}
y(l_1,M)y(l_2,M)y(l_3,M)
y(\tilde{l}_{123},M) F^2(l_1)F^2(l_2)F^2(l_3). 
\label{eqn:kurtinteg}
\end{eqnarray}
where $\tilde{l}_{123}=(\tilde{l}_{12}^2+l_3^2
-\sqrt{2}\tilde{l}_{12}l_3)^{1/2}$ with
$\tilde{l}_{12}=(l_1^2+l_2^2)^{1/2}$. 
Consequently, to obtain $\skaco{\kappa^4}^{1h}_c$, we need to perform a
5-dimensional numerical integration, which requires much less
computational time compared with the original 7-dimensional integration
of equation (\ref{eqn:conv4th}).
 
Figure \ref{fig:app_test} demonstrates the accuracy of our approximation
(\ref{eqn:appconv4th}). The approximate result for ${\cal I}(M,z)$ is
compared with the direct integration value, plotted against halo mass
$M$ for the $\Lambda$CDM model. The lens redshift is $z=0.4$ and
smoothing scale $\theta_{\rm s}=1'$.  Note that for fixed $z$ and $M$ we
can directly compute ${\cal I}(M,z)$ by evaluating the 5-dimensional
integral.  It is also worth noting that $z=0.4$ is chosen because it is
close to the peak of the lensing weight function $W(\chi,\chi_{\rm s})$
for source redshift $z_{\rm s}=1$.
The figure clearly shows that for both SCDM and \LCDM models our
approximation can reproduce ${\cal I}(M,z)$ within $10\%$ accuracy on
mass scales of $10^{14}M_\odot\simlt M\simlt 10^{16}M_\odot$, which
provide the dominant contributions to the kurtosis parameter on relevant
angular scales as shown in Figure \ref{fig:skurt_mass}.  The
approximation works better for more massive halos.

To estimate the final accuracy of our approximation to
$\skaco{\kappa^4}_c^{1h}$, we further need to take into account the lens
weighting, $W^4(\chi,\chi_{\rm s})d^{-6}_A$, as well as the weighting of
mass function, $f(\nu,z)M^3$, in equation (\ref{eqn:appconv4th}).  Since
the lens weighting gives a smooth redshift dependence, we here consider
the weighing of mass function.  The lower panel in Figure
\ref{fig:app_test} plots $f(\nu,z)M^3$ at $z=0.4$ against $M$, where
each curve is normalized to give unity at the peak mass scale.
Accounting for the weighting of $f(\nu)M^3$, we find that the accuracy
of our approximation is about $9\%$ and $8\%$ for SCDM and \LCDM models
at $\theta_{\rm s}=1'$ and $z=0.4$.
Our approximation works better for the larger smoothing scales, where
more massive halos contribute to $\skaco{\kappa^4}^{1h}_c$ (see Figure
\ref{fig:skurt_mass}).  From these results, we are confident that our
approximation can predict the 1-halo term within $\sim 10\%$ accuracy at
most on relevant scales, although we should bear in mind that the
approximation has a trend to underestimate the true value.  The figure
also shows the results from other possible approximations for \LCDM, as
in Figure \ref{fig:app_skew}, where we have used the replacements of
$y(l_{123})=y(l_1)y(l_2)y(l_3)$ (dotted line) and $y(l_{123})=y(l_3)$
(dot-dashed line) (Scoccimarro et al. 2001). These approximations
overestimate or underestimate ${\cal I}(M,z)$ by $62\%$ or $70\%$,
respectively, and become worse at more massive mass scales, and thus are
not accurate enough for our purpose.

Similarly, we can construct approximations for the 2-halo and 3-halo
terms to predict a total power of the convergence kurtosis.  The
explicit forms of the approximations used are presented in Appendix
\ref{conv23h}. We have confirmed that these approximations are
adequately accurate (see Scoccimarro et al. 2001 for similar discussions on
the skewness and kurtosis of the three-dimensional density field).  As
explained below, in this paper we ignore the 4-halo contribution which
is likely to have a negligible contribution on the angular scales we
have considered (see Cooray \& Hu 2001b for the trispectrum).

We can now compare our model predictions of the convergence kurtosis
parameter with the simulation results.  Figure \ref{fig:comp_kurt} plots
the result as in Figure \ref{fig:comp_skew}. It is apparent that our
halo model predictions are in good agreements with the simulation
results as for the skewness case. One caveat we should bear in mind is
again that the simulation result for \LCDM is likely to underestimate
$S_{\kappa,4}$ because of the reasons given for Figure
\ref{fig:comp_skew}. We have confirmed this by using new high-resolution
simulation data provided by Hamana (2002). We obtained
$S_{\kappa,4}\approx 4\times 10^4$ at $\theta_{\rm s}=1'$ for
$\Lambda$CDM, which gives a better match to our model prediction. The
main cosmological implication of this figure is that there are still
significant differences between SCDM and \LCDM models on small scales of
$\theta_{\rm s}\simlt 2'$, although the sampling errors corresponding to
a survey area of $25{\rm ~degree}^2$ become larger compared with the
skewness case.

The lower panel of Figure \ref{fig:comp_kurt} plots the 1-halo, 2-halo
and 3-halo contributions for $\Lambda$CDM. It is clear that the 1-halo
term gives the dominant contribution over the scales considered; the 2-halo
is marginally important on larger scales of $\theta_{\rm s}\simgt 5'$
and the 3-halo makes only a small contribution. More explicitly, these
terms provide $82\%$, $16\%$ and $2\%$ of the contributions to the total
kurtosis at $\theta_{\rm s}=1'$; $53\%$, $37\%$ and $10\%$ at
$\theta_{\rm s}=10'$, respectively. These results validate our
expectation that the 4-halo term is negligible on small angular scales
$\theta_{\rm s}\simlt 5'$, and even for larger scales it is likely to
have contributions smaller than $10\%$.  The 4-halo term is difficult to
evaluate by numerical integration because of the oscillatory shape of
the perturbation theory trispectrum, resulting from its dependences on
the interior angles of the 4-point configuration.

\section{Approximation for the shear kurtosis}
\label{shearapp}

Based on the results shown in the preceding section, we develop
an approximate method for calculating the shear kurtosis, which is the 
main purpose of this paper.

The connected fourth-order moment of the filtered shear field can be
expressed in terms of the convergence trispectrum as
\begin{eqnarray}
\skaco{\gamma_{1}^4(\theta_{\rm s})}_c=\int\!\!\prod_{i=1}^{4}
\frac{d^2\bm{l}_i}{(2\pi)^2}\cos2\phi_{l_i}F(l_i)
T_\kappa(\bm{l}_1,\bm{l}_2,\bm{l}_3,\bm{l}_4)(2\pi)^2\delta_D(\bm{l}_{1234}),
\label{eqn:gamma4th}
\end{eqnarray}
with 
\begin{equation}
T_\kappa\equiv \int\!\!d\chi W^4(\chi,\chi_{\rm s})d_A^{-6}T\!\left(
\bm{k}_1,\bm{k}_2,\bm{k}_3,\bm{k}_4\right),
\end{equation}
where $\bm{k}=\bm{l}/d_A(\chi)$, $\phi_{l_i}$ is defined by
$\bm{l}_i=l_i(\cos\phi_{l_i},\sin\phi_{l_i})$ and $T_\kappa$ is the
convergence trispectrum (see Cooray \& Hu 2001b).  This equation
clarifies that the integrand function of $\skaco{\gamma_1^4}_c$ has
configuration dependences via the geometrical factors of
$\cos2\phi_{l_i}$ in addition to the convergence trispectrum, and we
therefore have to consider the 8-dimensional integration.  Note that for
$\skaco{\gamma_2^4}_c$ the geometrical factor in equation
(\ref{eqn:gamma4th}) is $\prod_{i=1}^4\sin2\phi_{l_i}$, but
$\skaco{\gamma_1^4}_c=\skaco{\gamma_2^4}_c$ from statistical symmetry.
In comparing equation (\ref{eqn:gamma4th}) with equation
(\ref{eqn:conv4th}) for $\skaco{\kappa^4}_c$, the difference is only the
geometrical factor $\prod_{i=1}^{4}\cos2\phi_{l_i}$. We therefore expect
that the following simple relation
\begin{equation}
\skaco{\gamma_1^4(\theta_{\rm s})}_c\approx 
f_{g}\skaco{\kappa^4(\theta_{\rm s})}_c,
\label{eqn:facgeo}
\end{equation}
applies, with a constant factor $f_{g}$. We can derive an upper limit
for $f_g$ in the following rough manner. From the integrand function of
$\skaco{\gamma_1^4}_c$, we consider the angular averaged geometrical
function as a function of $l_1$, $l_2$ and $l_3$ defined by
\begin{equation}
{\cal G}(l_1,l_2,l_3)=\int\!\!\prod_{i=1}^3\frac{d\phi_{l_i}}{2\pi}
\cos2\phi_{l_i}\cos2\phi_{l_{123}}, 
\end{equation}
where $\cos\phi_{l_{123}}=-(l_1\cos\phi_{l_1}
+l_2\cos\phi_{l_2}+l_3\cos\phi_{l_3})/l_{123}$. The function ${\cal G}$
peaks at $l_1=l_2=l_3$ and approaches zero for $l_1\ll l_2,l_3$ or
$l_1\gg l_2,l_3$ and so on, so that the shear fourth-order moment is
suppressed compared to that of the convergence. Hence, the upper limit
on $f_g$ should be set by the case $l_1=l_2=l_3$: $f_g\le {\cal
G}(l,l,l)\approx 5.17\times 10^{-2}$.

Given the relation (\ref{eqn:facgeo}), the shear kurtosis can be
expressed in terms of the convergence kurtosis as
\begin{equation}
S_{\gamma,4}(\theta_{\rm s})
\equiv \frac{\skaco{\gamma_{i}^4}_c}{\sigma^6_{\gamma}(\theta_{\rm s})}
\approx 8f_{g}\frac{\skaco{\kappa^4(\theta_{\rm s })}_c}
{\sigma_\kappa^6(\theta_{\rm s})}
=8f_{ g}S_{\kappa,4}(\theta_{\rm s}),
\label{eqn:expratio}
\end{equation}
where $\sigma_{\gamma}(\theta_{\rm s})$ is the rms of the filtered shear
field defined by $\sigma_{\gamma}(\theta_{\rm s})\equiv
\skaco{\gamma_i^2(\theta_{\rm s})}^{1/2}$ and the factor of $8$ comes
from the relation $\skaco{\gamma_i^2}=\skaco{\kappa^2}/2$.  Note that
the upper limit on $f_g$ discussed above corresponds to $S_{\gamma,4}\le
0.41 S_{\kappa,4}$.

\begin{figure}
  \begin{center}
    \leavevmode\epsfxsize=8.4cm \epsfbox{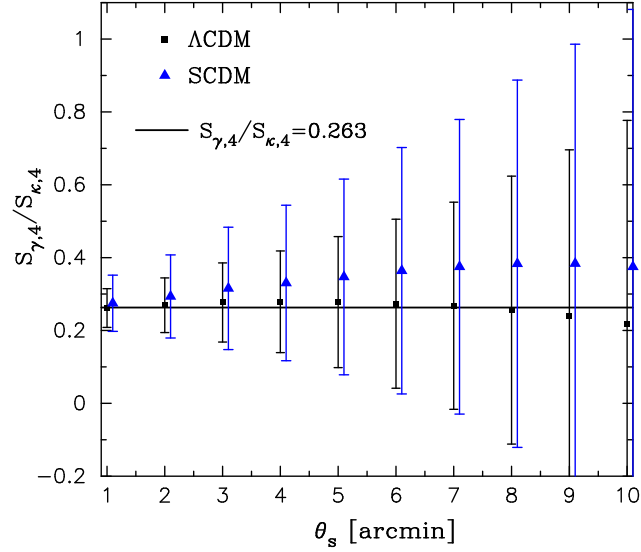}
  \end{center}
\caption{ Shown is the ratio of the shear kurtosis parameter to the
convergence kurtosis as a function of the smoothing scale.  The symbols with
error bars are the simulation results for the $\Lambda$CDM and SCDM
models as in Figure \ref{fig:comp_vari}, where the shear kurtosis is
obtained by averaging the two kurtosis parameters for $\gamma_1$ and
$\gamma_2$ and the error bar in each bin is properly averaged. The solid
line shows the ratio value of $0.263$, which is the average value
between the results at $\theta_{\rm s}=1'$ for the two models. }
\label{fig:ratio}
\end{figure}
Unfortunately, it is difficult to analytically derive the constant
factor connecting $S_{\gamma,4}$ and $S_{\kappa,4}$, and we therefore
rely on ray-tracing simulations.  Figure \ref{fig:ratio} plots the
simulation results for the ratio of $S_{\gamma,4}$ to $S_{\kappa,4}$ as
a function of the smoothing scale. Note that we have taken the average
of the kurtosis values for the two independent shear fields to obtain
$S_{\gamma,4}$; $S_{\gamma,4}\equiv
[\skaco{\gamma_1^4}_c/\skaco{\gamma_1^2}^3+
\skaco{\gamma_2^4}_c/\skaco{\gamma_2^2}^3]/2$. The figure reveals that,
despite the fact that the trispectrum has a strong dependence on
cosmological models (leading to a difference greater than $300\%$
between the kurtosis values for the SCDM and $\Lambda$CDM models -- see
Figure \ref{fig:comp_kurt}), the ratios are similar for the two
models. Further, it appears that $S_{\gamma,4}$ is related to
$S_{\kappa,4}$ by a constant factor over the angular scales we have
considered. The solid line shows the average value of $0.263$ between
the results at $\theta_{\rm s}=1'$ for the two models, and one can see
that the curve reasonably explains the simulation results. In the
following, to predict the shear kurtosis parameter, we simply multiply
the factor $0.263$ by the convergence kurtosis parameter calculated
using the approximations developed in \S \ref{appconv}. Thus we use:
\begin{equation}
S_{\gamma,4}\approx 0.263 \ S_{\kappa,4}. 
\label{eqn:ratio}
\end{equation}

\section{RESULTS}
\label{results} 

\begin{figure}
  \begin{center}
    \leavevmode\epsfxsize=8.4cm \epsfbox{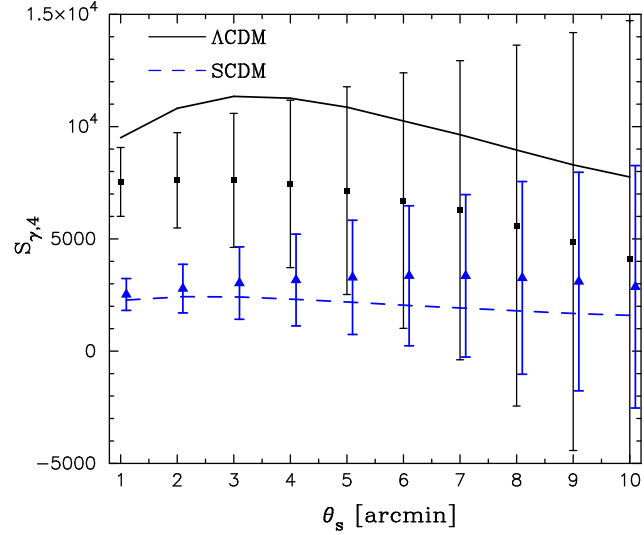}
  \end{center}
\caption{ Comparison of our model predictions for the shear kurtosis,
$S_{\gamma,4}$, with the simulation results as a function of the
smoothing scale. Here, for the simulation result in each bin we take the
average between the kurtosis parameters of two shear fields $\gamma_1$
and $\gamma_2$.  For illustration, we slightly shift the simulation
result for SCDM along the x-axis.  } \label{fig:comp_skurt}
\end{figure}
In Figure \ref{fig:comp_skurt} we compare our model predictions of the
shear kurtosis with the simulation results for the SCDM and \LCDM models
as in Figure \ref{fig:comp_kurt}. Note that the halo model prediction is
calculated from the sum of the 1-halo, 2-halo and 3-halo contributions
to the shear fourth-order moment.  The result shown in each bin is
computed from the average value of the kurtosis parameters for two shear
fields, $\gamma_1$ and $\gamma_2$, as in Figure \ref{fig:ratio}.  The
figure reveals that our model can well reproduce the simulation
results and that there are distinct differences between the shear
kurtosis values for the SCDM and $\Lambda$CDM models on small scales of
$\theta_{\rm s}\simlt 3'$. The range of angular scales with $0.5'\simlt
\theta_{\rm s}\simlt 3'$ is feasible for making adequate signal-to-noise
measurements of top-hat smoothed statistics from lensing survey data
(e.g., see Van Waerbeke et al. 2001a).
As mentioned in the discussion of Figure \ref{fig:comp_skew}, the
simulation results for \LCDM are likely to underestimate $S_{\gamma,4}$
at $\theta_{\rm s}\simlt 3'$. We have confirmed that high-resolution
simulation \cite{Hamana02} does give a better match to our model
prediction, but the accurate measurement of fourth-order statistics from
numerical data needs further investigation.  This uncertainty does not
seriously undermine our conclusions about parameter estimation from the
shear kurtosis, because it has a strong dependence on $\Omega_{\rm m0}$
for flat $\Lambda$CDM models. E.g. a small change $\Delta\Omega_{\rm
m0}=-0.05$ leads to a large change of $\Delta S_{\gamma,4}\approx
3.1\times 10^3$ at $\theta_{\rm s}=1'$ if one chooses the fiducial model
with $\Omega_{\rm m0}=0.3, \Omega_{\lambda0}=0.7$. On the other hand for
variations about the $\Omega_{\rm m0}=1$ model, the result is almost
unchanged with $\Delta S_{\gamma,4}\approx 60$. These scalings are
approximately consistent with the perturbation theory expectation given
by $S_{\gamma,4}\propto \Omega_{\rm m0}^{-2}$ (see also Figure
\ref{fig:skurt_cosmo} and \ref{fig:skurt_om}).

\begin{figure}
  \begin{center}
    \leavevmode\epsfxsize=8.cm \epsfbox{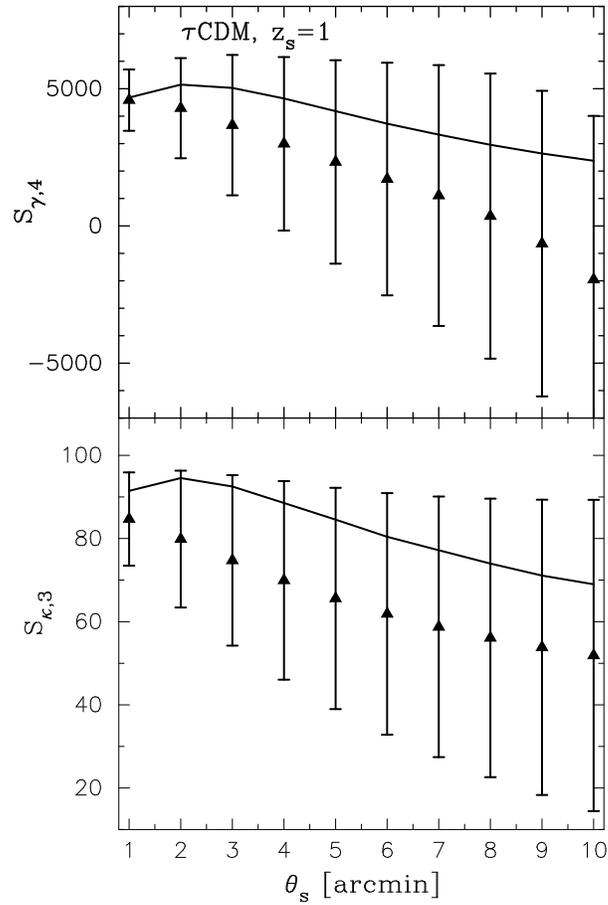}
  \end{center}
\caption{ Shown is the comparison of our model predictions for the shear
kurtosis (upper panel) and the convergence skewness (lower panel) with
the results of the JSW simulation data for the $\tau$CDM model, which
has $\Omega_{\rm m0}=1.0$, $h=0.5$, $\sigma_8=0.6$ and $\Gamma=0.21$.  }
\label{fig:tcdm}
\end{figure}
\begin{figure}
  \begin{center}
    \leavevmode\epsfxsize=8.cm \epsfbox{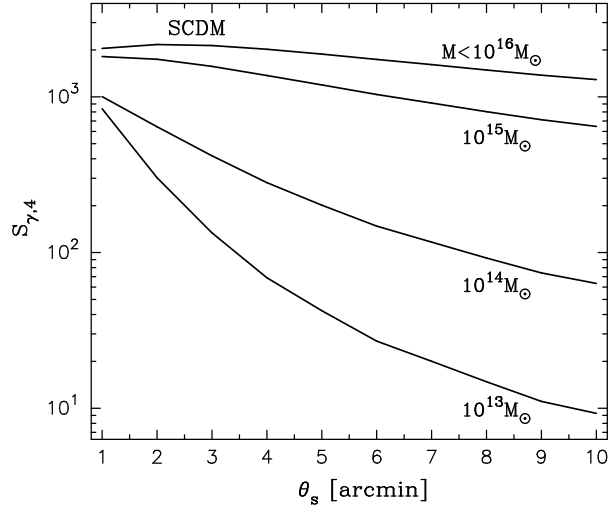}
  \end{center}
\caption{ The dependence of the shear kurtosis on the maximum mass
cutoff used in the calculation for the SCDM model.  Note the we varied
the maximum mass cutoff for evaluations of the shear fourth-order
moment and the shear variance, which enter the numerator and denominator
of $S_{\gamma,4}$, respectively.  From top to bottom, the maximum mass
is $10^{16}$, $10^{15}$, $10^{14}$ and $10^{13}M_\odot$.  }
\label{fig:skurt_mass}
\end{figure}
\begin{figure}
  \begin{center}
    \leavevmode\epsfxsize=8.cm \epsfbox{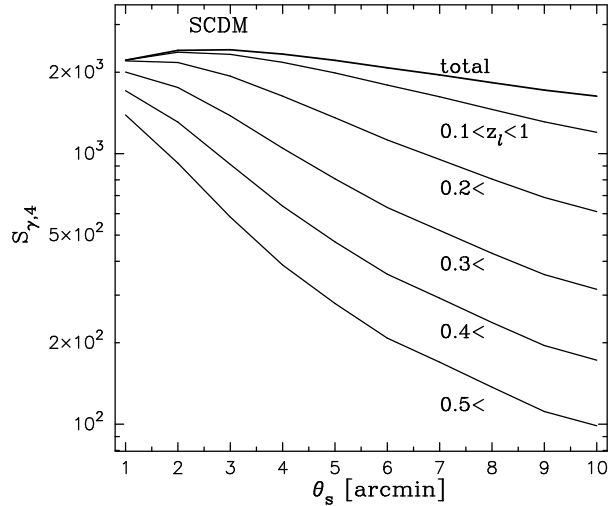}
  \end{center}
\caption{ The dependence of the shear kurtosis on the range of lens
redshift is plotted.  From top to bottom, the lower limit of lens
redshift used in the calculation is $0.1$, $0.2$, $0.3$, $0.4$ and
$0.5$, while the upper limit is $z=1$.  } \label{fig:skurt_lensz}
\end{figure}
\begin{figure}
  \begin{center}
    \leavevmode\epsfxsize=12.cm \epsfbox{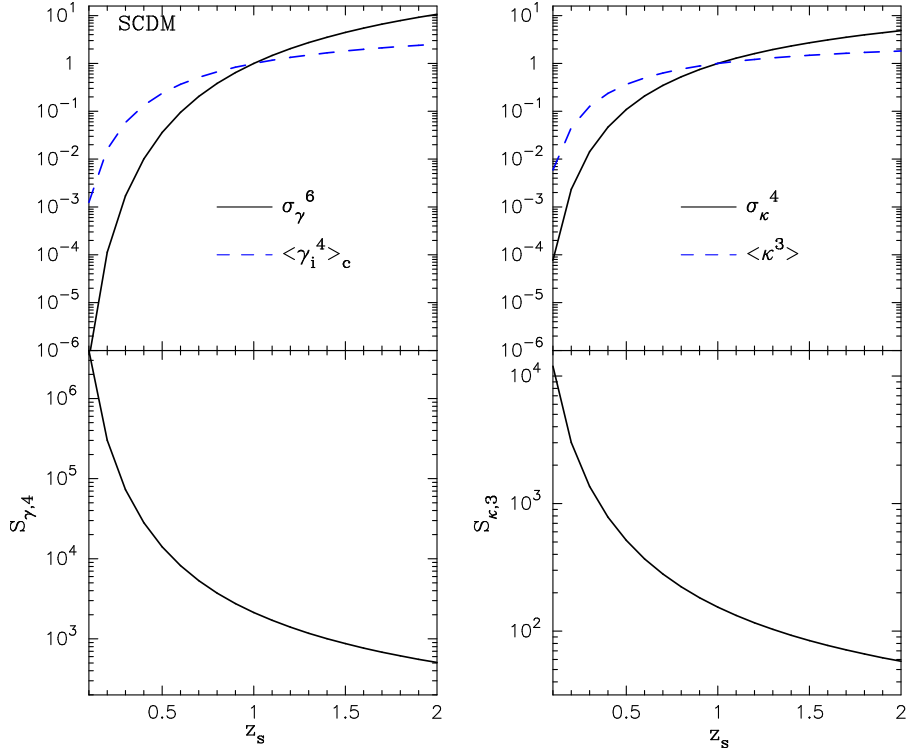}
  \end{center}
\caption{ In the left upper panel, the dependences of
$\sigma_\gamma^6(\theta_{\rm s})$ and the shear fourth-order moment on
the source galaxy redshift $z_{\rm s}$ are plotted for the SCDM model
and $\theta_{\rm s}=1'$, while the resulting dependence of the shear
kurtosis is shown in the lower left panel.  Note that $\sigma_\gamma^6$
appears in the denominator of $S_{\gamma,4}$.  The solid and dashed
lines in the upper panel are the results for $\sigma^6_\gamma$ and
$\skaco{\gamma_i^4}$, respectively, where each curve is normalized by
its values for $z_{\rm s}=1$.  The right panel is a similar plot for the
convergence skewness $S_{\kappa,3}$ given by equation
(\ref{eqn:skewconv}); $\sigma_\kappa^4$ and $\skaco{\kappa^3}$ are the
denominator and numerator of $S_{\kappa,3}$.}  \label{fig:skurt_zs}
\end{figure}
\begin{figure}
  \begin{center}
    \leavevmode\epsfxsize=8.cm \epsfbox{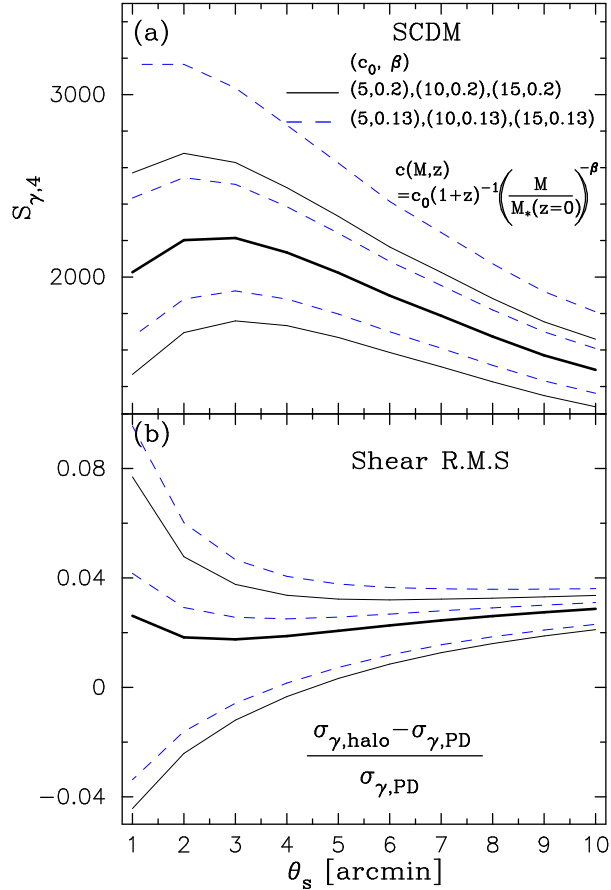}
  \end{center}
\caption{ In the upper panel (a), the dependence of the shear kurtosis
on variations in the concentration parameter is plotted as a function of
the smoothing scale.  We consider the concentration parameter expressed
in terms of the normalization $c_0$ at the present-day nonlinear mass
scale, $M_\star(z=0)$, and the slope of the mass dependence as
$c(M,z)=c_0(1+z)\left(M/M_\star(z=0)\right)^{-\beta}$.  The three solid
lines demonstrate the dependences for
$(c_0,\beta)=(5,0.2),(10,0.2),(15,0.2)$ from bottom to top,
respectively, while the dashed lines are
$(c_0,\beta)=(5,0.13),(10,0.13),(15,0.13)$. The bold solid line denotes
the result for $(c_0,\beta)=(10,0.2)$, which we have used in this paper.
In the lower panel (b), we show the relative differences between the
halo model predictions and the PD results for the shear rms. As in the
upper panel, the solid lines from bottom to top are the results for
$(c_0,\beta)=(5,0.2),(10,0.2),(15,0.2)$.  } \label{fig:skurt_conc}
\end{figure}
\begin{figure}
  \begin{center}
    \leavevmode\epsfxsize=8.cm \epsfbox{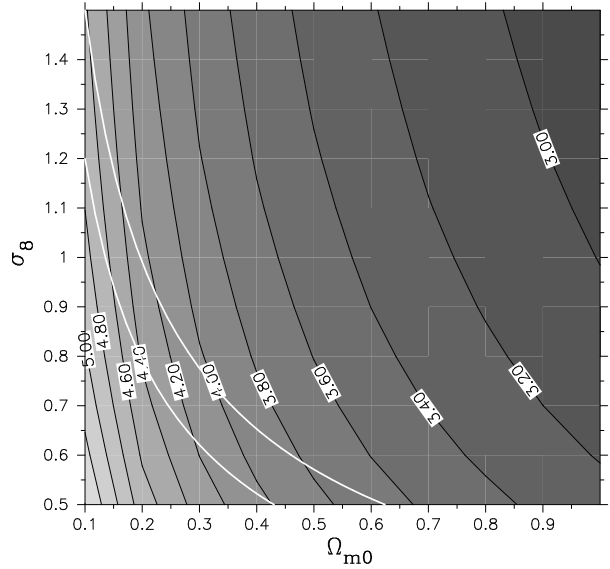}
  \end{center}
\caption{ Log contour map of the shear kurtosis parameter $S_{\gamma,4}$
in the $\Omega_{m0}$-$\sigma_8$ plane for flat CDM models with $h=0.7$
and $\theta_{\rm s}=1'$. Note that the number assigned to each contour
is the index value of $a$ which parameterizes the kurtosis as
$S_{\gamma}=10^a$; each contour is stepped by $\Delta a=0.2$.  The two
solid curves show the dependence $\sigma_8\Omega_{m0}^{0.6}$ that
represents typical constraints on the $\Omega_{\rm m0}$-$\sigma_8$ plane
so far obtained from the two-point shear statistics measurements (the
two curves correspond to two different normalizations).  It is clear
that the curves for the two-point and four-point statistics have
different shapes, and would therefore allow for independent
determinations of the two parameters. } \label{fig:skurt_cosmo}
\end{figure}
\begin{figure}
  \begin{center}
    \leavevmode\epsfxsize=8.cm \epsfbox{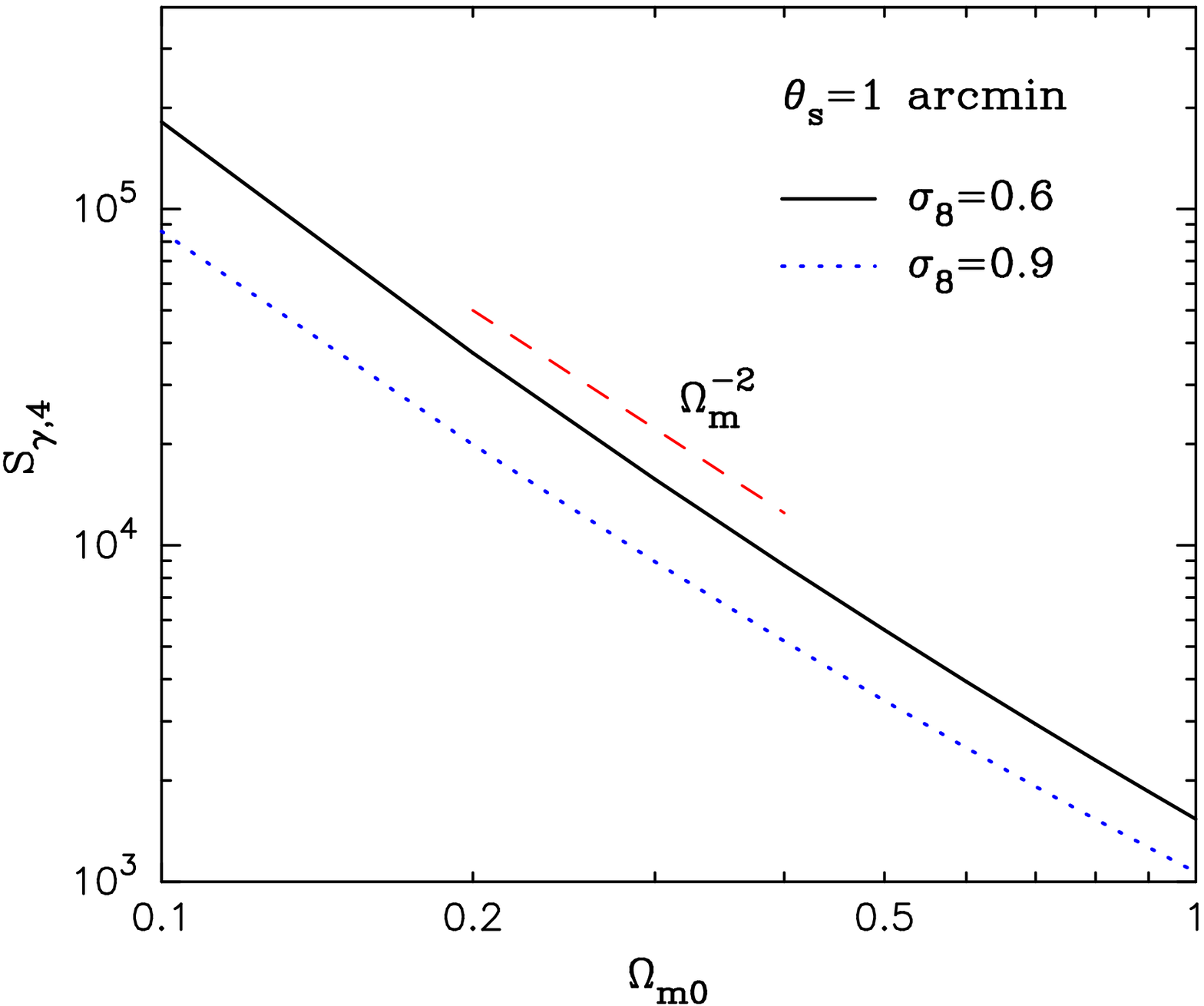}
  \end{center}
\caption{ This figure shows slices of the contour plot in Figure
\ref{fig:skurt_cosmo} with $\sigma_8=0.6$ and $0.9$, which explicitly
illustrates the dependence of the shear kurtosis on $\Omega_{\rm m0}$.
The dashed line shows the dependence $S_{\gamma,4}\propto \Omega_{\rm
m0}^{-2}$.}  \label{fig:skurt_om}
\end{figure}
In Figure \ref{fig:tcdm}, we show the comparison of our model prediction
with the JSW simulation results for the $\tau$CDM model, which has
$\Omega_{\rm m0}=1.0$, $h=0.5$, $\sigma_8=0.6$ and the shape parameter
of $\Gamma=0.21$ (see JSW for more details). Note that the cosmological
parameters of the $\tau$CDM model are the same as for SCDM except for
the shape parameter ($\Gamma=0.5$ for SCDM). The error in each bin is
the sample variance corresponding to a survey area of $\approx 7.84{\rm
~degree}^2$.  The purpose of this figure is to illustrate the validity
of our model for different cosmological models and the sensitivity of
the shear kurtosis to the shape of the matter power spectrum. One can
clearly see that our halo model reproduces the simulation results for
both the shear kurtosis (upper panel) and the convergence skewness
(lower panel). This success may be a surprise, because it has been
pointed out in several works (e.g. see Figure 1 in Van Waerbeke et
al. 2001b) that the JSW data does not exactly match the amplitude of the
convergence variance predicted by the PD formula.  The discrepancy may
be attributed to the inaccuracy of the PD fitting formula for the
nonlinear power spectrum. However it appears that our model can
reproduce the simulation results for the statistical measures of weak
lensing fields that are chosen to be insensitive to the power spectrum
normalization but can pick up the non-Gaussian signals originating from
the density field.  Figure \ref{fig:tcdm} also shows that the skewness
and kurtosis are larger for the $\tau$CDM model compared to the SCDM
model. Thus some constraint on the shape of the power spectrum is
necessary to use the non-Gaussian statistics for parameter estimation.

So far our halo model calculations have assumed the mass range for the
integration as $10^{3}M_\odot\le M\le10^{18}M_\odot$.  Figure
\ref{fig:skurt_mass} plots the dependence of the shear kurtosis on the
maximum mass cutoff used in the calculation for the SCDM model. Note
that we varied the maximum mass cutoff for evaluations of both the shear
variance and the fourth-order shear moment used in the calculation of
$S_{\gamma,4}$; the figure shows the resulting dependence of
$S_{\gamma,4}$ on the maximum mass cutoff.  It is apparent that
$S_{\gamma,4}$ is mainly due to massive halos with $M\simgt
10^{14}M_\odot$, while less massive halos contribute more on smaller
angular scales.  More specifically, halos with $M\ge 10^{14}M_\odot$
provide $\approx 50\%$ and $\approx 99\%$ contributions to the shear
kurtosis at $\theta_{\rm s}=1'$ and $10'$, respectively.

The dependence of the shear kurtosis on the lens redshift is shown in
Figure \ref{fig:skurt_lensz}.  This figure indicates that the shear
kurtosis is sensitive to low redshift structures with $z\simlt
0.4$. Next we examine the origin of the redshift dependence by plotting
dependences of the numerator and denominator of the skewness and
kurtosis separately.

The upper-left panel in Figure \ref{fig:skurt_zs} illustrates the
dependences of $\sigma_\gamma^6$ and $\skaco{\gamma_i^4}_c$ on the
source galaxy redshift $z_{\rm s}$ for the SCDM model and $\theta_{\rm
s}=1'$ ($\sigma_\gamma^6$ and $\skaco{\gamma_i^4}_c$ are the denominator
and numerator of $S_{\gamma,4}$, respectively).  For comparison, the
right panel shows a similar plot for the convergence skewness
$S_{\kappa,3}$ (the denominator and numerator of $S_{\kappa,3}$ are
$\sigma_\kappa^4$ and $\skaco{\kappa^3}$, respectively).  One can see
that nonlinear structures at lower redshifts affect the higher-order
moments more strongly than the terms with powers of the variance. The
lower-left panel plots the resulting dependence of $S_{\gamma,4}$ on the
source redshift and reveals that possible variation in the source
redshift alters the shear kurtosis.  A comparison of the left and right
panels shows that the shear kurtosis has a stronger dependence on the
source redshift than the convergence skewness.  The shear kurtosis
increases by a factor of 20 if the source redshift is varied from $0.5$
to $2$, while the convergence skewness varies about by a factor of 7.
These results raise the question: what is the best survey strategy to
measure the shear kurtosis?  Figure \ref{fig:skurt_zs} suggests that a
deeper redshift survey that probes higher-redshift structures loses some
non-Gaussian signal due to projection effects.  A survey to measure
$S_{\gamma,4}$ may be more efficient if it is shallower and covers
greater area, provided systematic errors are well understood and the
redshift distribution of source galaxies is known.
The feasibility of the measurement of non-Gaussian statistics
from lensing surveys will be presented in detail 
elsewhere (Takada et al. 2002).

In Figure \ref{fig:skurt_conc} we show the effect of varying the
concentration parameter of the NFW profile on the shear kurtosis for the
SCDM model and $\theta_{\rm s}=1'$. The dependence is illustrated by
parameterizing the concentration parameter in terms of its normalization
at the nonlinear mass scale today and the slope of the mass dependence
as $c(M,z)=c_0(1+z)^{-1}(M/M_\ast(z=0))^{-\beta}$. Here we have again
assumed that the redshift dependence is the same as in equation
(\ref{eqn:conc}) as suggested by the N-body simulations
\cite{Bullock01}. With fixed $\beta$, a $50\%$ increase or decrease of
$c_0$ leads to $\sim 30\%$ increase or decrease of $S_{\gamma,4}$. Thus
our results would not be strongly affected by varying $c$ to the extent
indicated by N-body simulations, which give a dispersion of 0.2 in $\ln
c$ (Jing 2000; Bullock et al. 2001; 
see also Cooray \& Hu 2001b for lensing study).  On the other
hand, the curves with fixed $c_0$ and varying $\beta$ reveal that the
shallower slope $\beta=0.13$ leads to a larger value of
$S_{\gamma,4}$. These results can be explained as follows. The increase
of $c_0$ for a given $\beta$ or the decrease of $\beta$ for a given
$c_0$ leads to more concentrated density profiles for halos more massive
than the nonlinear mass scale $M_\ast$. Since these massive halos dominate the
contribution to the shear kurtosis, this has the effect of increasing
the kurtosis on the angular scales considered here.
An important caveat is that the variations in the concentration
parameter simultaneously alter the predictions for the shear variance,
$\sigma^2_{\gamma}$. The lower panel shows the relative errors of the
halo model prediction to the PD results for $\sigma_\gamma$. It is clear
that our choice (\ref{eqn:conc}) for $c(M,z)$ (bold solid line) gives
the closest value to the PD result.  Furthermore, as we have shown, our
model can reproduce the simulation results for the higher-order moments
of weak lensing fields.  In this regard, therefore, as a prescription
for using the halo approach to study the higher-order statistics of weak
lensing, it is reasonable to choose the concentration parameter so that
it reproduces the PD result for the variance.  Conversely, if we employ
a different halo profile from NFW, it will probably be necessary to
modify our choice (\ref{eqn:conc}) for the concentration parameter in
order to accurately describe the higher-order statistics.

Finally, in Figure \ref{fig:skurt_cosmo} we show the contour plot for
$S_{\gamma,4}$ with $\theta_{\rm s}=1'$ in the plane of $\Omega_{\rm
m0}$ and $\sigma_8$ parameters for flat CDM models with $h=0.7$. The
number assigned to each contour denotes the value of
$\log_{10}S_{\gamma,4}$, in the $\Omega_{\rm m0}-\sigma_8$
plane. Clearly, the shear kurtosis is very sensitive to $\Omega_{\rm
m0}$ and has a weak dependence on $\sigma_8$.  For example, the model
with $\Omega_{\rm m0}=0.1$ and $\sigma_8=0.6$ yields
$S_{\gamma,4}=1.72\times 10^5$, while the model with $\Omega_{\rm
m0}=1.0$ and the same $\sigma_8$ leads to $S_{\gamma,4}=1.46\times
10^3$.  Figure \ref{fig:skurt_om} shows slices of the contour plot with
$\sigma_8=0.6$ and $0.9$, and reveals that the dependence of
$S_{\gamma,4}$ on $\Omega_{\rm m0}$ is very close to
$S_{\gamma,4}\propto \Omega_{\rm m0}^{-2}$.  On the other hand, the two
solid lines in Figure \ref{fig:skurt_cosmo} show the dependence
$\sigma_8\Omega_{\rm m0}^{0.6}$, which represents typical constraints
obtained from measurements of the two-point statistics of the shear
field (each curve is arbitrarily normalized). One can see that these two
curves have a very different shape from the contours of $S_{\gamma,4}$;
thus measurements of $S_{\gamma,4}$ can break the degeneracy in the
determination of $\Omega_{\rm m0}$ and $\sigma_8$. Furthermore, they can
constrain the dark energy component of the universe if they are combined
with the evidence for a flat universe from recent CMB measurements
\cite{Netter01}.

\section{DISCUSSION AND CONCLUSION}
\label{disc}

In this paper we have investigated the kurtosis parameter of the cosmic
shear field, $S_{\gamma,4}(\equiv
\skaco{\gamma_i^4}_c/\skaco{\gamma_i^2}^3)$, based on the dark matter
halo approach.  The two main results revealed in this paper are
summarized as follows. First, we have developed a useful approximation
for calculating the shear kurtosis, which significantly reduces the
computational time and yet provides the shear kurtosis expected within
$\sim 10\%$ accuracy over the angular scales $1'\le\theta_{\rm
s}\le 10'$. Our model predictions can well
match the ray-tracing simulation results for the shear kurtosis as well
as for the convergence skewness and kurtosis parameters 
for the SCDM, $\Lambda$CDM and $\tau$CDM models (see Figure
\ref{fig:comp_skew}, \ref{fig:comp_kurt}, \ref{fig:comp_skurt}
and\ref{fig:tcdm}). For the \LCDM model, the simulation data lie 
slightly below our predictions. It appears that the numerical results
on small scales, especially for the higher-order moments, have not
converged at the few percent level of accuracy -- this is a subject
that merits further investigation. While we have focused on lensing
statistics in this paper, our results for the higher-order moments
apply to the 3-dimensional density field as well. We show in Figure 6
that our approximations allow for the 3-dimensional skewness to be
accurately computed down to sub-Mpc scales, which is an improvement
on existing approaches in the literature. 

Second, we have shown that $S_{\gamma,4}$ has a strong dependence on the
matter density parameter of the universe, $\Omega_{\rm m0}$, while it is
only weakly dependent on the power spectrum normalization, $\sigma_8$, as
illustrated in Figure \ref{fig:skurt_cosmo} and
\ref{fig:skurt_om}. Thanks to this property, a measurement of $S_{\gamma,4}$, 
in combination with the shear two-point statistics already measured, 
would be valuable in constraining both  $\Omega_{\rm m0}$ and the matter
power spectrum. For example, a marginal
detection of the shear kurtosis with $50\%$ uncertainties would
yield the constraint $0.24\simlt \Omega_{\rm m0}\simlt0.43$ if the
current concordance model with $\Omega_{\rm m0}=0.3$,
$\Omega_{\lambda0}=0.7$, $h=0.7$ and $\sigma_8=0.9$ is taken as the
fiducial model. 
Even a null detection of $S_{\gamma,4}$ allows us to set a lower limit on
$\Omega_{\rm m0}$ from the strong dependence of $S_{\gamma,4}$ 
on low $\Omega_{\rm m0}$ values.  Thus
measurements of $S_{\gamma,4}$ can break the degeneracies in the
$\Omega_{\rm m0}$ and $\sigma_8$ determination so far provided from the
shear two-point statistics measurements without invoking any other
methods.  It can
determine the dark energy component of the universe if
combined with the strong evidence of a flat universe from the CMB data. 
It could also help resolve
the puzzling inconsistency in the determination of $\sigma_8$ in the `old'
(e.g., Eke et al. 1996) and 'new' (e.g., Seljak 2001) cluster abundance
estimations (see also Van Waerbeke et al. 2002 and Lahav et al. 2001
for comments on this issue from analyses of the weak lensing and the
galaxy redshift survey, respectively).  

We believe that the shear kurtosis is more
directly applicable to data from weak lensing surveys than the
well-studied higher-order statistics of the convergence field. To 
examine this issue in detail, we must examine the signal-to-noise
properties of different measures of non-Gaussianity from realistic
survey data. This would facilitate a comparison of different approaches, 
such as the shear kurtosis discussed here and the shear 3-point function
proposed by Bernardeau et al (2002a). 

\begin{figure}
  \begin{center}
    \leavevmode\epsfxsize=9.cm \epsfbox{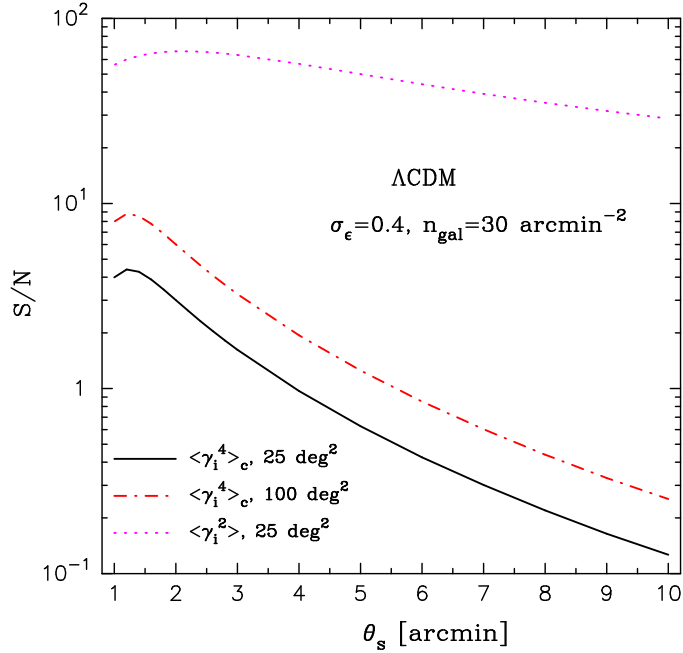}
  \end{center}
\caption{ Estimates of the signal-to-noise ratio in the
measurement of the connected fourth-order moment of the shear field,
$\skaco{\gamma_i^4}_c$, for the \LCDM model against the
smoothing scale. We assume $\sigma_\epsilon=0.4$ and $n_{\rm gal}=30
~{\rm arcmin}^{-2}$ for the rms intrinsic ellipticity and the number
density of source galaxies, and take the survey area to be $\Omega_{\rm
survey}=25~{\rm degree}^2$. 
For comparison, the dotted and dot-dashed lines are the
results for the shear variance and $\skaco{\gamma_i^4}_c$ with
$\Omega_{\rm survey}=100~{\rm degree}^2$, respectively. }
\label{fig:sn}
\end{figure}
\begin{figure}
  \begin{center}
    \leavevmode\epsfxsize=8.cm \epsfbox{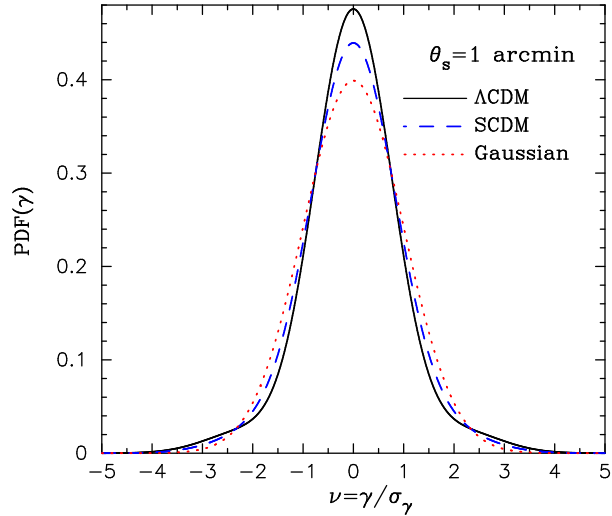}
  \end{center}
\caption{ The probability distribution function (PDF) of the shear
field.  The solid and dashed lines are PDFs for the LCDM and SCDM models
and $\theta_{\rm s}=1'$, which are calculated using the Edgeworth
expansion (\ref{eqn:edge}), while the dotted line is the Gaussian PDF.
}  \label{fig:edge}
\end{figure}
For this purpose, it is crucial to correctly model possible
errors in measurements of the shear, since 
higher-order statistics can be very sensitive to the noise. 
The main sources of error are
the shot noise due to the intrinsic ellipticities of source galaxies and
the sampling error for a finite survey area. 
If the
intrinsic ellipticity distribution is regarded as Gaussian owing to random
intrinsic orientations, we can estimate the dispersion for a measurement
of the connected fourth-order moment of the shear field,
$\skaco{\gamma_i^4}_c$, following the method developed in \S 5 in
Schneider et al. (1998):
\begin{equation}
 \sigma(\skaco{\gamma_i^4}_c)\approx 
\frac{\theta_{\rm s}} {\sqrt{\Omega_{\rm survey}}}
\sqrt{5451\times \sigma_\gamma^8(\theta_{\rm s})+\frac{24}
{(\pi \theta_{\rm s}^2 n_{\rm gal})^4}
\left(\frac{\sigma_{\epsilon}^2}{2}\right)^4},
\label{eqn:sn}
\end{equation}
where $\sigma_\epsilon$ is the dispersion of the intrinsic ellipticity
distribution, $\Omega_{\rm survey}$ the survey area and $n_{\rm gal}$
the number density of source galaxies. The first term on the r.h.s of
the equation above denotes the sample variance and the second term the
noise due to the finite number of randomly located source galaxy
images. Here we have assumed $n_{\rm gal}\theta_{\rm s}^2\gg 1$ and that
the connected part of the higher-order moments of the shear field is
equal to its unconnected part for simplicity \footnote{For the
fourth-order moment, the ratio of the connected part to the unconnected
part, $\skaco{\gamma_i^4}_c/[3\skaco{\gamma_i^2}^2]$, is less than $0.5$ 
on the angular scales we have considered. If this also holds for
the sixth- and eighth-order moments, equation (\ref{eqn:sn}) would
give a conservative estimate of the noise.}.  Figure \ref{fig:sn} shows an
estimate of the signal-to-noise ratio in the measurement of
$\skaco{\gamma_i^4}_c$ for the \LCDM model at $z_{\rm s}=1$, where we
have assumed $n_{\rm
gal}=30~{\rm arcmin}^{-2}$ and $\sigma_\epsilon=0.4$ and considered two
cases of $\Omega_{\rm survey}=25$ and $100~{\rm deg}^2$. Note that a
signal-to-noise ratio for the measurement of $S_{\gamma,4}$ is similar
for the result in this figure, since the error arises mainly from the
measurement of $\skaco{\gamma_i^4}_c$ compared with that of
$\skaco{\gamma_i^2}$, as shown by comparing the solid and dotted lines.  The
noise is mainly due to the sample variance at $\theta_{\rm s}\simgt
2'$ for the \LCDM model, while the intrinsic ellipticity noise is
important at $\theta_{\rm s}\simlt 2'$.  One can see that for a survey
area of 25 square degrees the measurement of $\skaco{\gamma_i^4}_c$
would indeed be marginally feasible on small angular scales $\theta_{\rm
s}\simlt 3'$, provided systematic errors can be kept under
control. The results also imply an interesting possibility as discussed
in Figure \ref{fig:skurt_zs}: a shallower survey, for a given amount
of observing time, could improve the
signal-to-noise ratio because the amplitude of $\skaco{\gamma_i^4}_c$
does not decrease as much as that of the variance for low redshift
structures. Further the redshift distribution is easier to measure 
accurately for a shallower survey. In any case, with a survey 
area exceeding 100 square
degrees, expected from forthcoming lensing surveys, the kurtosis 
measurement should be made with high statistical significance, as shown by the
dot-dashed line, and thus prove useful for parameter estimation.

There are some uncertainties we have ignored in the rough signal-to-noise
estimate of equation (\ref{eqn:sn}).  
First, non-Gaussian errors are more important on
smaller angular scales \cite{CH01b}, so sample variance must be 
estimated by using an adequate number of realizations of
ray-tracing simulations or possibly by an analytical treatment using the
halo approach for calculating the connected sixth- and eighth-order moments. 
Second, in actual data the noise distribution
of the intrinsic ellipticities is likely to be non-Gaussian.
In addition, we should bear in mind that the shear kurtosis is strongly
affected by rare events in the tail of the measured shear
distribution. Hence, we will need to consider some strategy to
efficiently extract the shear kurtosis from realistic data which is
less sensitive to unphysical rare events. One possible way to reduce
the sample variance from such rare events is to use the probability
distribution function (PDF) of the shear field, which is analogous to
the method proposed by JSW for the study of the convergence skewness
parameter. If the primordial fluctuations are Gaussian, the nonlinear
gravitational evolution of structure formation induces the
non-Gaussianity in the weak lensing fields, as investigated in this
paper, and the weakly non-Gaussian PDF of the shear field can be modeled
by the Edgeworth expansion (see Juszkiewicz et al. 1995).  At the lowest
order, we have
\begin{eqnarray}
P(\gamma_i)=\frac{1}{\sqrt{2\pi}\sigma_{\gamma}(\theta_{\rm s})}
\exp\left(-\frac{\gamma_i^2}{2\sigma_{\gamma}^2(\theta_{\rm s})}\right)
\left[1+\frac{1}{4{\rm !}}S_{\gamma,4}(\theta_{\rm s})
\sigma_{\gamma}^2(\theta_{\rm s}) H_4\left(\frac{\gamma_i}
{\sigma_{\gamma}(\theta_{\rm s})}\right)\right],
\label{eqn:edge}
\end{eqnarray}
where $H_4(x)=x^4-6x^2+3$ is the fourth-order Hermite polynomial.  The
resulting PDFs for the SCDM and \LCDM models are shown in Figure
\ref{fig:edge}.  Moreover, in practice we must 
account for the fact that the measured shear field is a sum of the
cosmic shear and noise fields. To obtain the PDF for the measured shear
field $\gamma_i^{\rm obs}$, therefore, we have to convolve $P(\gamma_i)$
with the PDF of the noise, where the noise field is defined by
{\em smoothing} the intrinsic ellipticities of source galaxies
contained within top-hat apertures.  Note that there
are two noise fields ($\epsilon_1$ and $\epsilon_2$) corresponding
to the two shear fields ($\gamma_1$ and $\gamma_2$), respectively.  If
the noise PDF is given by $P_N(\epsilon_i)$, the PDF for the measured
shear field, $\gamma_i^{\rm obs}$, can be expressed by the convolution
integral as
\begin{eqnarray}
P(\gamma^{\rm obs}_i)&=&
\int\!\!d\epsilon \int\!\!
d\gamma_i P_N(\epsilon_i)P(\gamma_i)
\delta_D(\gamma_i^{\rm obs}-\gamma_i-\epsilon),
\label{eqn:pdf}
\end{eqnarray}
where $P_N(\epsilon_i)$ is normalized as
$\int\!\!d\epsilon_iP_N(\epsilon_i)=1$. It is worth noting that the noise PDF
$P_N(\epsilon_i)$ can  also be modeled in terms of the variance and
higher-order moments of the noise field using the Edgeworth expansion.
Furthermore, this method can utilize a great advantage -- the noise PDF
$P_N(\epsilon_i)$ can be directly reconstructed from the observed shear
field, e.g. by smoothing after the randomization of the position angle of 
each galaxy image. 
This procedure would wash out the {\em coherent} cosmic shear pattern within
the smoothing aperture, but pick up the contribution from the intrinsic
ellipticities of source galaxies.  
In this sense, the variance and higher-order moments of
the noise field for a given smoothing scale can be directly extracted
from the measured data, giving an estimator for the noise PDF. 
To obtain the noise for
the convergence field is harder, since a non-local
reconstruction from the smoothed intrinsic ellipticity field is required
\cite{vW99a}.  
For the shear, the theoretical model (\ref{eqn:pdf}) for $P(\gamma^{\rm
obs}_i)$ is given by a single parameter,
$S_{\gamma,4}(\theta_{\rm s})$. One can then fit the theoretical
prediction to the measured PDF over an appropriate intermediate range of
$\gamma^{\rm obs}_i$, where the Edgeworth expansion is valid, 
to extract the shear kurtosis $S_{\gamma,4}$ with reduced
sensitivity to rare events. The quantitative improvement in the 
signal-to-noise will be the subject of a later study. 

Other uncertainties we have ignored in this paper are effects of
the redshift distribution of source galaxies and the clustering of source
galaxies. Figure \ref{fig:skurt_zs} shows that the higher-order
moments are more sensitive to nonlinear structures at lower redshifts
than the variance and, as a result, the shear kurtosis can be
sensitive to the source redshift distribution.  Consider the
conventionally used model for the redshift distribution of source galaxies
\begin{equation}
n(z)=\frac{\beta}{z_0\Gamma(\frac{1+\alpha}{\beta})}
\left(\frac{z}{z_0}\right)^\alpha\exp
\left[-\left(\frac{z}{z_0}\right)^\beta\right],
\end{equation}
with $\alpha=2$ and $\beta=1.2$ (e.g., see van Waerbeke et al. (2002))
and the source redshift parameter $z_0=0.48$, so that it
gives the mean source redshift $\skaco{z_{s}}\approx 1.0$. 
This distribution increases the value of $S_{\gamma,4}$ at $\theta_{\rm
s}=1'$ by $\sim 20\%$ for the SCDM and \LCDM models, compared to 
the case with source redshifts fixed at $z_s=1$.  This increase is
to some extent counterbalanced by the source clustering effect, because
previous work based on perturbation theory \cite{Bern98} showed that
source clustering reduces the values of the convergence skewness and kurtosis
by $\sim 10-20\%$ (recently confirmed by ray-tracing
simulation \cite{HamanaSource} as well as by an analytical study
\cite{Schneider01}).  It is reasonable to expect that this argument is
also valid for the shear kurtosis, because we showed $S_{\gamma,4}$ is
related to the convergence kurtosis $S_{\kappa,4}$ through a 
geometrical constant factor (see discussions around equation
(\ref{eqn:ratio})).
 
Although the halo approach used here and in the literature assumes a
spherically symmetric profile, in reality halos have 
non-spherical profiles and substructure as predicted in the
CDM paradigm (e.g., Jing \& Suto 2001).  
Our results showed that the halo model can fairly reproduce the
ray-tracing simulation results for the 1-point moments
of the smoothed lensing fields. This success is
encouraging, since the simulations include contributions from various
realistic halo profiles.  The agreement is partly because we focus only on 
statistical quantities and, therefore, to some extent 
the profile we need for the halo model calculation should be 
an average over possible halo profiles of a given mass. In
addition, 1-point moments of the smoothed fields are likely to be
insensitive to halo profile fluctuations.  We expect that the
full 3- or 4-point correlation functions of the lensing fields would be 
more sensitive to profile fluctuations, because those functions
should contain complete information on gravitational clustering up to
the 3 or 4-point level through the configuration dependences. These issues
will be presented elsewhere (Takada \& Jain 2002).

Finally, we comment on an alternative application of measurements of the shear
kurtosis.  If cosmological parameters including $\Omega_{\rm m0}$ are 
precisely determined by other measurements, 
our results suggest that the higher-order
moments of weak lensing fields could be used to constrain 
dark matter halo profiles. This
problem is particularly interesting, since it can be a clue to
understanding the nature of dark matter.  Although the concentration and
inner profile of dark matter halo are degenerate in giving two-point
lensing statistics as argued in this paper (see also Seljak
2000), a detailed study may yield ways of combining the two-point and 
four-point  shear statistics
to break the degeneracy by exploiting the dependences
of the shear kurtosis on the inner profile and the concentration 
parameter shown in Figure \ref{fig:skurt_conc}.  If this is the
case, Figure \ref{fig:skurt_mass} indicates that measurements of the shear
kurtosis at $\theta_{\rm s}\simgt 1'$ can constrain the properties
of the halo profile at mass scales $M\simgt 10^{14}M_\odot$.

\bigskip

We would like to thank T. Hamana for kindly providing his ray-tracing
simulation data as well as for fruitful comments. This work benefitted
from several discussions with R. Scoccimarro.  We also thank
E. Komatsu, A. Taruya and L. van Waerbeke for valuable
discussions and D. Dolney for a careful reading of the manuscript.  
This work is supported in part by the 
Japan Society for Promotion of Science (JSPS)
Research Fellowships, a NASA-LTSA grant, and a Keck foundation grant.


\appendix
\section{Perturbation theory bispectrum and trispectrum}
\label{app:pert}

The explicit forms of the bispectrum and trispectrum of the density field
based on perturbation theory (e.g., Fry 1984) are 
\begin{eqnarray}
B^{\rm pt}(\bm{k}_1,\bm{k}_2,\bm{k}_3,z)&=&2F_2(\bm{k}_1,\bm{k}_2)
P^L(k_1,z)P^L(k_2,z)+ \mbox{2 perm.},\\
T^{\rm pt}(\bm{k}_1,\bm{k}_2,\bm{k}_3,\bm{k}_4; z)&=&4\left[
F_2(\bm{k}_{13},-\bm{k}_1)F_2(\bm{k}_{13},\bm{k}_2)P^L(k_{13},z)
P^L(k_1,z)P^L(k_2,z)
+ \mbox{11 perm.} \right]\nonumber\\
&&+6\left[F_3(\bm{k}_1,\bm{k}_2,\bm{k}_3)+
P^L(k_1,z)P^L(k_2,z)P^L(k_3,z)+
\mbox{3 term}\right]\label{eqn:perttrisp},
\end{eqnarray}
where the redshift evolution of the linear power spectrum is given by
$P^L(k,z)=D^2(z)P^L(k,z=0)$ where $D(z)$ is the growth factor.  
The kernels
$F_n$ are calculated using perturbation theory (e.g., see Jain \&
Bertschniger 1994) and are expressed as
\begin{eqnarray}
F_2(\bm{k}_1,\bm{k}_2)&=& \frac{5}{7}+\frac{1}{2}\left(\frac{1}{k_1^2}+
\frac{1}{k_2^2}\right)(\bm{k}_1\cdot\bm{k}_2)
+\frac{2}{7}\frac{(\bm{k}_1\cdot\bm{k}_2)^2}{k_1^2k_2^2},\nonumber\\
F_3(\bm{k}_1,\bm{k}_2,\bm{k}_3)&=&\frac{7}{18}\frac{\bm{k}_{12}
\cdot\bm{k}_1}{k_1^2} 
\left[F_2(\bm{k}_2,\bm{k}_3)+G_2(\bm{k}_1,\bm{k}_2)\right]+
\frac{1}{18}\frac{k_{12}^2(\bm{k}_1\cdot\bm{k}_2)}{k_1^2k_2^2}
\left[G_2(\bm{k}_2,\bm{k}_3)+G_2(\bm{k}_1,\bm{k}_2)\right],
\end{eqnarray}
with
\begin{equation}
G_2(\bm{k}_1,\bm{k}_2)=\frac{3}{7}+\frac{1}{2}\left(\frac{1}{k_2^2}
+\frac{1}{k^2_1}\right)\frac{\bm{k}_1\cdot\bm{k}_2}{k_1k_2}+
\frac{4}{7}\left(\frac{\bm{k}_1\cdot\bm{k}_2}{k_1k_2}\right)^2,
\end{equation}
where we have ignored the extremely weak dependences of the
functions $F_n$ and $G_2$ on cosmological parameters $\Omega_{\rm m0}$
and $\Omega_{\lambda 0}$.

\section{Geometrical properties of top-hat filter function}
\label{app:tophat}

The purpose of this Appendix is to derive properties of the 
integrals of products of the two-dimensional top-hat kernel given by
equation (\ref{eqn:tophat}). This is 
analogous to the approach in Appendix B of Bernardeau (1994).

We begin our discussion with deriving the following identity for the
third-order products, since it is relevant for the calculation of the
third-order moment of weak lensing fields (see \S \ref{app}):
\begin{equation}
\int\!\!\frac{d^2\bm{x}_1}{(2\pi)^2}\int\!\!\frac{d^2\bm{x}_2}{(2\pi)^2}
F(x_1)F(x_2)F(x_{12})=\int_0^\infty\!\!\frac{x_1dx_1}{2\pi}
\int_0^\infty\!\!\frac{x_2dx_2}{2\pi}F^2(x_1)F^2(x_2), 
\label{eqn:formtop}
\end{equation}
where the top-hat kernel is $F(x)=2J_1(x)/x$ and $x_{12}\equiv |\bm{x}_1+\bm{x}_2|$.  

The proof of equation (\ref{eqn:formtop}) is as follows.  From the
expansion formula of Bessel function (e.g., 8.532.1 in Gradshteyn \&
Ryzhik 2000) we can expand the kernel $F(|\bm{x}_1+\bm{x}_2|)$ as
\begin{eqnarray}
F(|\bm{x}_1+\bm{x}_2|)=2\frac{J_1(|\bm{x}_1+\bm{x}_2|)}{|\bm{x}_1+\bm{x}_2|}&=&4
\sum_{n=0}(n+1)\frac{J_{n+1}(x_1)}{x_1}\frac{J_{n+1}(x_2)}{x_2}
(-1)^{n}\frac{\sin[(n+1)\Phi_2]}{\sin\Phi_2},
\end{eqnarray}
where $\Phi_2$ is the angle between $\bm{x}_1$ and $\bm{x}_2$, giving
$x_{12}=(x_{1}^2+x_2^2-2x_1x_2\cos\Phi_2)^{1/2}$. Inserting this
equation into the left part of equation (\ref{eqn:formtop}) and then
integrating it over the angle $\Phi_2$, we obtain
\begin{equation}
\int\!\!\frac{x_1dx_1}{2\pi}\frac{x_2dx_2}{2\pi}F(x_1)F(x_2)
\left[F(x_1)F(x_2)-12\frac{J_3(x_1)}{x_1}\frac{J_2(x_2)}{x_2}
+20\frac{J_5(x_1)}{x_1}\frac{J_5(x_2)}{x_2}+\cdots\right],
\label{eqn:expbes}
\end{equation}
Using the recursion relation for Bessel functions and an integration 
formula (6.574.2 in Gradshteyn \& Ryzhik 2000), the second term in
the bracket on the r.h.s of equation above vanishes because 
\begin{eqnarray}
\int^\infty_0\!\!\frac{x_1dx_1}{2\pi}\frac{J_1(x_1)}{x_1}\frac{J_3(x_1)}{x_1}
&=&\int^\infty_0\!\!\frac{x_1dx_1}{2\pi}\frac{J_1(x_1)}{x_1}
\left[4\frac{J_2(x_1)}{x_1^2}-\frac{J_1(x_1)}{x_1}\right]\nonumber\\
&=&4\cdot 2^{-1}\frac{\Gamma(2)\Gamma(1)}{\Gamma(3/2)\Gamma(5/2)\Gamma(1/2)}
-\frac{\Gamma(1)\Gamma(1/2)}{\Gamma(1/2)\Gamma(3/2)\Gamma(1/2)}\nonumber\\
&=&\frac{1}{2}-\frac{1}{2}=0. 
\end{eqnarray}
One can similarly find that the third term and higher terms in the
bracket on the r.h.s of equation (\ref{eqn:expbes}) vanish, and then
equation (\ref{eqn:formtop}) follows.

Likewise, one can straightforwardly obtain the following identity on
which the approximation for calculating the convergence forth-order
moment is based:
\begin{equation}
\int\!\!\prod_{i=1}^3\frac{d^2\bm{x}_i}{(2\pi)^2}
F(x_1)F(x_2)F(x_3)F(x_{123})
=\int\!\!\prod_{i=1}^3\frac{x_idx_i}{2\pi}
F^2(x_1)F^2(x_2)F^2(x_3), 
\end{equation}
where $x_{123}\equiv |\bm{x}_1+\bm{x}_2+\bm{x}_3|$. 

\section{2-halo and 3-halo contributions to the convergence
 fourth-order moment} \label{conv23h}

In this Appendix, we write down the expressions for approximations used
for calculations of the 2-halo and 3-halo contributions to the
convergence fourth-order moment, which are discussed in \S \ref{app}.

The 2-halo term of the convergence fourth-order moment receives two
contributions, which represent taking three or two particles
in the first halo:
\begin{equation}
\skaco{\kappa^4}^{2h}_c=\int\!\!d\chi W^4(\chi,\chi_{\rm s})d_A^{-4}(\chi)
\int\!\!\!\prod_{i=1}^3\frac{d^2\bm{l}_i}{(2\pi)^2}\left[T^{2h}_{31}+T^{2h}_{22}
\right](\bm{l}_1,\bm{l}_2,\bm{l}_3,-\bm{l}_{123})F(l_i)F(l_{123}),
\end{equation}
where $T^{2h}_{31}$ and $T^{2h}_{22}$ are given by equations
(\ref{eqn:trisp2h31}) and (\ref{eqn:trisp2h22}), 
and $\bm{k}_i=\bm{l}_i/d_A(\chi)$.  It is clear that
the contribution from $T^{2h}_{31}$ dominates that from $T^{2h}_{22}$,
because the former arises mainly from 3-point correlations within one
halo with a highly nonlinear density contrast, while the latter
arises from 2-point correlations.  We confirmed that the $T^{2h}_{22}$
contribution is smaller even than the 3-halo term. For this reason, we
ignore the $T^{2h}_{22}$ contribution and use the following
approximation for calculating the 2-halo term in the convergence
fourth-order moment (we also include the approximation for 
$F(l_{123})$ from equation (\ref{eqn:apptop})):

\begin{eqnarray}
\skaco{\kappa^4}^{2h}_c&\approx&4\int\!\!d\chi W^4(\chi,\chi_{\rm s})
d_A^{-6}(\chi)
\int\!\!dM_1\frac{dn}{dM_1}\left(\frac{M_1}{\bar{\rho}_{0}}\right)^3
b(M_1,z)\int\!\!\frac{l_1dl_1}{2\pi}\frac{l_2dl_2}{2\pi}y(l_1,M_1)y(l_2,M_1)
y(\tilde{l}_{12},M_1)F^2(l_1)F^2(l_2)\nonumber\\
&&\times \int\!\!dM_2\frac{dn}{dM_2}b(M_2,z)\left(\frac{M_2}{\bar{\rho}_0}\right)
\int\!\!\frac{l_3dl_3}{2\pi}y(l_3,M_3)F^2(l_3)P^L(l_3),
\label{eqn:app2h}
\end{eqnarray}
where $\tilde{l}_{12}=(l_1^2+l_2^2-l_1l_2)^{1/2}$ and the factor of $4$
comes from the permutation symmetry in the 2-halo trispectrum (see
equation (\ref{eqn:trisp2h31})) for the fourth-order moment
calculation. We have written the order of integration specifically to point
out that the last two integrals, over $M_2$ and $l_3$, can be performed
separately from the preceding three. Thus one needs to perform 
at most a 4-dimensional numerical integration to get $\skaco{\kappa^4}^{2h}_c$.

Similarly, we use the following equation to calculate the 3-halo term in
the convergence fourth-order moment: 
\begin{eqnarray}
\skaco{\kappa^4}^{3h}_c&\approx&6\int\!\!d\chi W^4(\chi,\chi_{\rm s})
d_A^{-6}(\chi)\int\!\!\frac{l_1dl_1}{2\pi}\frac{d^2\bm{l}_2}{(2\pi)^2}
B^{\rm pt}(\bm{l}_1,\bm{l}_2,-\bm{l}_{12})F^2(l_1)F^2(l_2)
\int\!\!dM_1\frac{dn}{dM_1}\frac{M_1}{\bar{\rho}_0}
b(M_1,z)y(l_1,M_1)\nonumber\\
&&\times \int\!\!dM_2\frac{dn}{dM_2}\frac{M_2}{\bar{\rho}_0}
b(M_2,z)y(l_2,M_2)\int\!\!dM_3\frac{dn}{dM_3}\left(\frac{M_3}{\bar{\rho}_0}\right)^2
b(M_3,z)\int\!\!\frac{l_3dl_3}{2\pi} y^2(l_3,M_3)F^2(l_3),
\label{eqn:app3h}
\end{eqnarray}
where $d^{2}\bm{l}_2=l_2ddl_2d\Phi_2$,
$l_{12}=(l_1+l_2-2l_1l_2\cos\Phi_2)^{1/2}$ and the factor of 6 comes
from equation (\ref{eqn:trisp3h}).  We have used the approximation
$y(l_{123})=y(l_3)$, which is valid to high accuracy for the 3-halo
term. Again, a careful consideration of the dependences of the
integrated functions reveals that we have to perform only a
4-dimensional numerical integration to obtain $\skaco{\kappa^4}^{3h}_c$.


\label{lastpage}
\end{document}